
\input harvmac.tex

\def\frac#1#2{{#1\over#2}}

\def\exp{{\rm exp}}

\def\slash#1{\mathord{\mathpalette\c@ncel{#1}}}
\overfullrule=0pt

\def\steepslash{\c@ncel}
\def\frac#1#2{{#1\over #2}}

\def\p {\partial}

\def\C{{\bf C}}
\def\inbar{\,\vrule height1.5ex width.4pt depth0pt}
\def\IB{\relax{\rm I\kern-.18em B}}
\def\IC{\relax\hbox{$\inbar\kern-.3em{\rm C}$}}
\def\IP{\relax{\rm I\kern-.18em P}}
\def\IR{\relax{\rm I\kern-.18em R}}
\def\IZ{\relax\ifmmode\mathchoice
{\hbox{Z\kern-.4em Z}}{\hbox{Z\kern-.4em Z}}
{\lower.9pt\hbox{Z\kern-.4em Z}}
{\lower1.2pt\hbox{Z\kern-.4em Z}}\else{Z\kern-.4em Z}\fi}

\catcode`\@=12
\def\tr{{\rm tr}}

\Title{\vbox{\baselineskip12pt\hbox{RU-93-30}
                \hbox{hept-th/9307196}}}
{\vbox{\centerline{Free Field Representation}
\vskip6pt\centerline{For Massive Integrable Models}}}

\centerline{Sergei Lukyanov \footnote{$\dagger$}
{On leave of absence from L.D. Landau Institute for Theoretical
Physics, Kosygina 2, Moscow, Russia}
\footnote{$^*$}{e-mail address: sergei@physics.rutgers.edu}}
\bigskip\centerline{Department of Physics and Astronomy}
\centerline{Rutgers University, Piscataway, NJ 08855-049}
\centerline{}
\centerline{}

\centerline{\bf{Abstract}}
A new  approach to massive integrable models is
considered.
It allows  one to find
symmetry algebras which define the   spaces of local
operators and to get  general  integral representations
for form-factors  in  the\ $ SU(2)$\   Thirring
and Sine-Gordon models.

\Date{July, 93}

\newsec{Introduction}

Two dimensional integrable field theory today is among
the most advanced topics
in relativistic field theory. The reason essentially lies
in specific two-dimensional symmetries which
lead to exact solutions of the quantum field dynamics.

In a
massive theory these symmetries
show up as  a  drastically simplified scattering
theory called Factorized Scattering Theory (FST).
This structure  was first observed in non relativistic scattering
of spin waves
\ref\bet{H. A.Bethe, Z.Phys. 71 (1931) 205.}
and quantum particles with point-like interaction
\ref\ber{F.A. Berezin, J.P. Pokhil  and  V.M. Finkelberg,
Vestn. Mosk. Univ. Mat. Mekh., 1 (1964) 21.},
\ref\gui{J.B.  Mc. Guire, J. Math. Phys. 5 (1964) 622 .},
and also
in classical scattering of solitons in nonlinear field models
\ref\gard{C.S.Gardner, T.M. Green, M.D.Kruskal and R.M. Miura,
Phys.Rev.Lett. 19 (1967) 1095.},
\ref\zakh{V.E. Zakharov, ZhETF 60 (1970) 993.}.
Factorized scattering  preserves the number of particles
and the set of their on-mass-shell momenta. This conservation is ensured by
an infinite series of commuting integrals of
motion
\ref\ar{I.Ya. Arefyeva  and V.E. Korepin, Pisma ZhETF 21 (1974) 160.},
\ref\kul{P.P. Kulish, Theor. Math. Phys. 26 (1976) 188.}.
The computation of the exact factorized S-matrix may be
performed by combining the standard requirements of unitarity and crossing
symmetry together with the symmetry properties of the model
\ref\z{A.B. Zamolodchikov, Commun. Math. Phys. 55 (1977) 183},
\ref\shr{B. Shroer, T.T. Truong and P.H. Weisz, Phys. Lett.
B67 (1977) 321.},
\ref\za{A.B. Zamolodchikov and Al.B. Zamolodchikov,
Ann. Phys. (NY) 120 (1979) 165.}.
Large variety of the factorized scattering theories was
constructed explicitly (see e.g.
\ref\og{E.I. Ogievetski, N.Yu. Reshetikhin and P.B. Wiegmann,
Nucl. Phys. B280 [FS 18] (1987), 45.},
\ref\fat{V.A. Fateev and A.B. Zamolodchikov,
"Conformal Field Theory and Purely Elastic S-Matrices", in
"Physics and Mathematics of Strings",
memorial volume for Vadim Knizhnik, L. Brink, D. Friedan,
A. Polyakov eds., World Scientific (1989).},
\ref\car{J. Cardy and G. Mussardo, Phys. Lett.
B229 (1989) 243.},
\ref\fre{P. Freund, T. Classen and E. Melzer,  Phys. Lett. 229B (1989) 243;
G. Sotkov and C. Zhu, Phys. Lett. 229B (1989) 391;  P. Christe  and
G. Mussardo, Nucl. Phys. B330 (1990) 465.} ).

The two particle S-matrix  uniquely specifies a  structure of a
space of local operators for integrable models .
In other words, its knowledge can be used to compute
off-shell quantities, like correlation functions of elementary or composite
fields of the  integrable models under investigation. This can be
achieved by considering the form-factors of local fields, which
are matrix elements of operators between asymptotic states
\ref\kar{M. Karowski and P. Weisz, Nucl. Phys. B139 (1978) 455.},
\ref\berg{B. Berg, M. Karowski and P. Weisz, Phys. Rev. D19 (1979) 2477.}.
A very important step in this direction
was taken in a series of papers
\ref\smi{F.A. Smirnov, Journal Phys. A 17 (1984) L873.},
\ref\sm{F.A. Smirnov, Journal Phys. A 19 (1986) L575.},
\ref\ki{A.N. Kirillov and F.A. Smirnov, Phys. Lett. B198 (1987) 506.},
where it  was shown that general properties of
unitarity, analyticity and locality lead, in the
case of  factorized scattering,
to a system of functional equations on form-factors which is
powerful enough to permit   the  reconstruction
of the matrix elements
(see also
\ref\smirn{F.A. Smirnov, Form-Factors in Completely Integrable Models of
Quantum Field Theory, World Scientific (1992) .},
\ref\yur{V.P. Yurov and Al.B. Zamolodchikov,
Int. Journ. Mod. Phys. A6 (1991) 3419.}).

In this article I discuss a  method
to describe the  space of\   local operators for massive integrable models.
It can be considered as a generalization of the  Feigin-Fuchs
representation in two dimensional Conformal Field Theory (CFT) .
The  bosonization in CFT allows one
to express   local operators
in terms of simple boson fields.
Using well known  examples of
FST, I shall demonstrate  that  the Smirnov equations
can
be rewritten as
a set of requirements   for  some  boson field.
The properties of its\  two point function
naturally generalize the properties of free boson Green function in CFT.
Methods of the bosonization permit one
to find symmetry   algebras defining the structure
of the space of local operators and
to get general  integral representations for form-factors
in the $SU(2)$ Thirring  and Sine-Gordon models.

The paper is organized as follows.

Sec. 2
contains  well known facts from FST.
For specialists it can be useful only as a list of
necessary notations.
Here I introduce the central  object of the investigation:\ the formal
Zamolodchikov-Faddeev algebra
\za ,
\ref\fad{L.D. Faddeev, Sov. Sci. Rev. Math. Phys. , 1C (1980) 107.}.
The Hilbert space of a  massive
integrable model is  a space of  representation of this algebra.
Such representations will be  denote   as \ $\pi _A$\ .
General ideas of this work are illustrated by the
\ $SU(2)$ invariant Thirring model\  ($SU(2)$\ TM)
\ref\pea{P.K. Mitter and P.H. Weisz, Phys. Rev. D8 (1973) 4410.}
\ref\bank{T. Banks, D. Horn and H. Neuberger, Nucl.Phys B108 (1976) 119. },
\ref\hal{M.B. Halpern, Phys. Rev. D12 (1976) 1684.}
and the Sine-Gordon model (SGM)
\ref\col{M. Ablowitz, P. Kaup, A. Newell and H. Segur, Phys. Rev. Lett. 30
(1973) 1262.},
\ref\ztt{V.E. Zakharov, L.A. Takhtadjyan and
L.D. Faddeev, Sov. Phys. Dokl. 19
(1974/75).}.
So I recall   essential properties of
these theories. At the end of the
section,  the definition of form-factors is
introduced and     the  Smirnov equations are formulated.

In the next section,  I argue  the main idea of the work.
{}From the  mathematical point of view, the
Smirnov equations are a kind of   Riemann-Hilbert problem.
Its solution is based on the following observation; \
Form-factors can be expressed as some traces
over the proper representation of the  Zamolodchikov-Faddeev
algebra. I shall denote
these representations as\ $\pi_Z$\ . They  essentially
differ from representations\ $\pi_A$\ . One can regard \ $\pi_Z$\
as a space of angular quantization of an integrable model
\ref\zamolodc{A.B. Zamolodchikov, unpublished.}.
I should note that a similar idea was used for solving the
quantum Knizhnik-Zamolodchikov equation
\ref\fr{I.B. Frenkel and N.Yu. Reshetikhin, Commun. Math.
Phys. 146 (1992) 1.}.
This is not surprising since the   Smirnov  and  quantum
Knizhnik-Zamolodchikov equations   have  close structures.
Moreover, they
are equivalent for some cases
\ref\sm{F.A. Smirnov, Dynamical symmetries of
massive integrable models I,II, RIMS preprints 772;
838 (1991).}.
Closely similar techniques  have been used  in
a remarkable series of works  on lattice models
\ref\japone{O. Foda and T. Miwa, Int. J. Mod. Phys. A7 Suppl.
1A (1992) 279.}, \ref\japtwo{B. Davies, O. Foda, M. Jimbo, T. Miwa
and A.Nakayashiki, Comm. Math. Phys. 151 (1993) 67.}
\ref\japthree{M. Jimbo, K. Miki, T. Miwa and A. Nakayashiki,
Phys. Lett. A168 (1992) 256.},
\ref\newjap{M. Jimbo, T.  Miwa and A.  Nakayashiki,
Difference equations for the correlation functions
of the eight-vertex model,
RIMS preprint 904 (1992).}.

In Sec. 4-8  the formal
constructions from Sec. 3 are illustrated  by  \ $SU(2)$\ TM.
I discus this model in detail  as
the simplest example where the  general  bosonization technique can be applied.
In this case the  construction
is equivalent to the  Frenkel-Jing bosonization
of the affine quantum algebra\ $U_q(\widehat{ sl(2)})$\
\ref\fren{I.B. Frenkel and N.H. Jing, Proc. Nat'l. Acad. Sci. USA
85 (1988) 9373.}.
In the Sec. 6 the symmetry algebra of the  space of local operators
is  found.
Then using the well known  method  from String Theory
\ \ref\cls{L. Clavelli and J.A. Shapiro, Nucl. Phys. B57 (1973) 490.},
I  get
the integral representation
for   form-factors generating functions.
One can expect that
they form a general
solution of the   Smirnov equations for  \ $SU(2)$\ TM.

In the Section 9 the  method  is applied
to SGM.
I  show that
the Zamolodchikov-Faddeev algebra for SGM admits
the free boson representation.
The classical limit of this representation was first obtained
in the work
\ref\shat{S. Lukyanov and S. Shatashvili, Phys. Lett. B298 (1993) 111.}.
The bosonization for  SGM
is the
natural  generalization of the Feigin-Fuchs representation in 2D CFT
\ref\fei{B.L. Feigin and D.B. Fuchs, unpublished (1983).},
\ref\dot{Vl.S. Dotsenko and V.A. Fateev, Nucl. Phys. B240 (1984) 312;
B251 (1985) 691.}.
I believe also that it has the same
base as the  bosonizations
of  the quantum affine algebra\ $U_q(\widehat{ sl(2)})$\
for general level
\ref\mat{A. Matsuo, Free Field Representation of
Quantum Affine Algebra \ $U_q(\widehat{sl(2)})$,
Nagoya University preprints, August 1992.}
\ref\japnew1{J. Shiraishi,
Free boson representation of $U_q(\widehat{sl(2)})$,
Tokyo preprint, UT-617 (1992); A. Kato, Y. Quano and
J. Shiraishi, Free boson representation
of q-vertex operators and their
correlation functions, Tokyo preprint UT-618 (1992)},
\ref\bour{A. Abada,  A.H. Bougourzi and M.A. El Gradechi,
Deformation of the Wakimoto construction, Montreal preprint CRM-1829 (1992).}.
As  in the $SU(2)$\ TM the  bosonization
permits one  to get the integral representation  for  form-factors of
SGM.

Some details of the  construction are described in the
Appendices.

The main technical results of this work are represented by
the Theorem   and Proposition 4 in
Sec.7,   together with the explicit  bosonization rules .

\newsec{Preliminaries}

\subsec{ Zamolodchikov-Faddeev algebra}
We start with a  brief description of the basics  of FST.

The massive character
of a spectrum  means that
the interaction in a theory is short-distance and, asymptotically for
$t\to {\pm}\infty$, the particles behave as free ones. So, two
natural different bases  can be
chosen for the Hilbert space of a  massive
field theory: initial (in) and final (out) bases of states.
Any in- and out-state
\eqn\sas{\eqalign{|A_{a_1}(p_1)...A_{a_n}(p_n)>_{in}, \ & \
|A_{a_1}(p_1)...A_{a_n}(p_n)>_{out}; \cr
 p^1_1<p^1_2&<...<p^1_n\cr}}        is characterized by a set
of particles $\{A_a\}$ and their two- momenta $p^{\mu}_a\ (\mu=x,t)$ :
\eqn\as{(p_a^t)^2-(p_a^x)^2=m_a^2.}
Here index $a$ denotes some
quantum numbers  specifying  different types of particles\ $A_a$\
with the masses\ $m_a$\ .
The in- and
out-bases should be connected by a proper
unitary matrix, which is called S-matrix
\eqn\sm{|A_{a_1}(p'_1)...A_{a_n}(p'_n)>_{out}
=S^{b_1...b_n}_{a_1...a_n}(p'_1,...p'_n|p_1,...p_n)
A_{b_1}(p_1)...A_{b_n}(p_n)>_{in}.}
Since the dynamics of integrable models is governed by an infinite number of
nontrivial conservation laws the scattering processes
are purely elastic and a
general n-particle element of  the S-matrix is
factorizable into  the two-particle
S-matrices
\eqn\sma{|A_{a_1}(p_1)A_{a_2}(p_2)>_{out}=
S^{b_1b_2}_{a_1a_2}(p_1,p_2)|A_{b_1}(p_1)A_{b_2}(p_2)>_{in}.}
It is convenient to parameterize the energy-momentum spectrum \as \   in terms
of  the
rapidity variable $\beta$
\eqn\be{p^0_a=m_a\cosh\beta, \ \  p^1_a=m_a\sinh\beta.}
By Lorentz invariance, the two particle scattering
amplitude will be a function of the
rapidity difference\   $\beta=\beta_1-\beta_2$\ only.

What can be said about the matrix function \ $S^{b_1b_2}_{a_1a_2}(\beta)$\ ?
In order to avoid some technical complication
we will consider
FST which contains only particles of the same mass
in the  spectrum\foot{We assume that there are no bound states.}.
One can suppose that they are
arranged in  a  multiplet of a  some finite dimensional (quantum) group\ $G$\ .
In this case  general principles of Quantum Field Theory and
factorization condition lead to the following requirements for the
two-particle \ $S$-matrix\ \z\ ,\shr ,\za :

1. The matrix function $S^{b_1b_2}_{a_1a_2}(\beta)$ must be
analytic
in the physical strip\  $0\leq\Im m\beta\leq\pi $\ .

2. Unitarity condition
\eqn\uni{S^{b_1b_2}_{a_1a_2}(\beta)S^{c_1c_2}_{b_1b_2}(-\beta)
=\delta^{c_1}_{a_1}\delta^{c_2}_{a_2}.}

3. Crossing symmetry
\eqn\cros{S^{b_1b_2}_{a_1a_2}(i\pi-\beta)=
\C_{a_1c}S^{cb_2}_{da_2}(\beta)\C^{db_1}.}
Here\ $\C_{ab}$ is the charge conjugation matrix and
  $$\C_{ab}\C^{bc}=\delta^c_a.$$

4. Yang-Baxter equation:
\eqn\yb{\eqalign{S^{c_1c_2}_{a_1a_2}(\beta_1-\beta_2)
S^{b_1c_3}_{c_1a_3}(&\beta_1-\beta_3)S^{b_2b_3}_{c_2c_3}(\beta_2-\beta_3)
=\cr
&=S^{c_1b_3}_{a_1c_3}(\beta_1-\beta_3)
S^{c_2c_3}_{a_2a_3}(\beta_2-\beta_3)
S^{b_1b_2}_{c_1c_2}(\beta_1-\beta_2).\cr}}

To describe   a space of states  in a  massive integrable model, it is
convenient to introduce the formal Zamolodchikov-Faddeev
algebra\ \za ,
\fad  . It is generated by the operators
\ $V_a(\beta)$\  which satisfy
the commutation
relation
\eqn\fz{V_{a_1}(\beta_1)V_{a_2}(\beta_2)
=S^{b_1b_2}_{a_1a_2}(\beta_1-\beta_2)V_{b_2}(\beta_2)V_{b_1}(\beta_1).}
Asymptotic states \ \sas\ form  the space of
representation of the
Zamolodchikov-Faddeev algebra.
We will denote it  as
\ $\pi_A$\ and
$$A_a(\beta)=\pi_A [V_a(\beta)].$$
One can interpret  \ $A_a(\beta)$\  as
particle creation operators, so
\eqn\hrt{\eqalign{&|A_{a_1}(p_1)...A_{a_n}(p_n)>_{out}=
A_{a_1}(\beta_1)...A_{a_n}(\beta_n)|vac>,\cr
&|A_{a_1}(p_1)...A_{a_n}(p_n)>_{in}=
A_{a_n}(\beta_n)...A_{a_1}(\beta_1)|vac>,}}
where \ $|vac>$\ is the vacuum state (the state without any particle)
and \ $\beta_1<\beta_2<...<\beta_n$ .
The conjugate operators
annihilate the vacuum state
\eqn\retsxv{\bar A^a(\beta)|vac>=0.}
They satisfy
the commutation relations:
\eqn\erwe{\bar A^{a_1}(\beta_1) A_{a_2}(\beta_2)=
A_{b_2}(\beta_2) S^{a_1b_2}_{b_1a_2}(\beta_2-\beta_1)\bar A^{b_1}(\beta_1)+
2\pi \delta^{a_1}_{a_2}\delta(\beta_1-\beta_2).}
The equation\ \erwe\ specifies
the structure of the   Hilbert space on \ $\pi_A$\ .

Another important  class of operators acting in the space \ $\pi_A$\
is
an  infinite set \ $\{I_s\}$\  of local commutative
integrals of motion (IM)
\ar,\ \kul .
The index \ $s$\ denotes the  spin of the conserved charge \ $I_s$.
The asymptotic states \ \hrt\ diagonalise  local IM
\eqn\bcv{\eqalign{&I_s|vac>=0\ ,\cr
I_s|A_{a_1}(p_1)...A_{a_n}(p_n)>_{in,out}&=
\gamma^{(s)}\
\sum_{k=1}^{n}
\exp(\beta_k s)|A_{a_1}(p_1)...A_{a_n}(p_n)>_{in,out},}}
where \ $\gamma^{(s)}$\ are some numbers.
It is convenient to consider the generating function\ $I(\alpha)$\  such that
\eqn\ikjyh{\eqalign{&I(\alpha)=\sum_{s>0}I_s\exp(-\alpha s),\ \
\exp(\alpha)\to \infty\ ,\cr
&I(\alpha)=\sum_{s>0}I_{-s}\exp(\alpha s),\ \
\exp(\alpha)\to \ 0 .}}
The formula\ \bcv\ means that the  generating function \  \ikjyh\
obeys  the following commutation relation with operators\
$A_a(\beta)$:
\eqn\gdfsd{\big[I(\alpha),A_a(\beta)\big]=\p_{\alpha}
\ln s(\alpha-\beta) A_a(\beta).}
The scalar function\ $s(\alpha)$\ is an important characteristic of an
integrable model.

Before ending this subsection, I  wish to point out that the
formal Zamolodchikov-Faddeev algebra\ \fz\
may  have other interesting
types of  representations when relation \ \erwe\ are not satisfied and
the operators \ $ \pi[V_a(\beta)]$\
do  not admit  the such simple physical meaning as
 \ $A_a(\beta)$\ \japtwo\ .

\subsec
{The Factorized Scattering Theories
for  \ $ SU(2)$\
TM  and SGM}
Now, we shall consider examples of FST which contain
only two particles \ $A_a\ (a=\pm1)$\  in their spectrum.

Suppose that the formal Zamolodchikov-Faddeev algebra\ \fz\  is
defined by the two particle S-matrix with the following non
trivial elements\ \za ,
\ref\berg{B. Berg, M. Karowski, V. Kurak and P. Weisz,
Nucl. Phys. B134 (1978) 125.},
\eqn\jvhbf{\eqalign{&S_{++}^{++}(\beta)=S_{--}^{--}(\beta)=S(\beta),\cr
&S_{+-}^{+-}(\beta)=S^{-+}_{-+} (\beta)=S(\beta)\frac{\beta}{i \pi-\beta},\cr
&S_{+-}^{-+}(\beta)=S^{+-}_{-+}(\beta)=S(\beta)\frac{i\pi}{i \pi-\beta},}}
here
\eqn\jfhv{S(\beta)=\frac{\Gamma(\frac{1}{2}-\frac{i\beta}{2 \pi})
\Gamma(\frac{i\beta}{2 \pi})}{\Gamma(\frac{1}{2}+\frac{i\beta}{2 \pi})
\Gamma(-\frac{i\beta}{2 \pi})}.}
The S-matrix\ \jvhbf\  satisfies  all the  axioms of  FST,
if the charge
conjugation matrix\ $\C_{ab}$\ is given by
\eqn\char{\C_{ab}=\delta_{a+b,0}.}

One can define
an action of the Lie algebra \ $sl(2)$\
on the   Hilbert space \ $\pi_A$\ corresponding to\  \jvhbf\
as follows
\ref\lus{M. Luscher, Nucl. Phys. B135 (1978) 1.},
\ref\lec{D. Bernard and A. LeClair, Comm. Math. Phys. 142 (1991) 99.};

1.We  assume that the vacuum state\ $|vac>$\ is
a \ $sl(2)$\ singlet:
\eqn\vb{Q^{\pm}|vac>=Q^{0}|vac>=0,}
where
$$Q^{\pm}=\pi_A[X^{\pm}]\ ,Q^{0}=\pi_A[H]$$
and \ $\{X^{\pm},H\}$\ is the Cartan-Weyl basis of \ $sl(2)$\ :
\eqn\tte{\eqalign{&[H,X^{\pm}]=\pm \sqrt2 X^{\pm},\cr
&[X^{+},X^{-}]=\sqrt2  H.}}

2.
One particle states\ $|A_a(\beta)>$\ are identified  with the fundamental
representation\ $ {\cal V}$\  of the Lie algebra\ $sl(2)$\ :
\eqn\kjh{\eqalign{&Q^{0}|A_a(\beta)>=\frac{a}{\sqrt2}|A_a(\beta)>,\cr
&Q^{\pm}|A_{\pm}(\beta)>=0,\cr
&Q^{\pm}|A_{\mp}(\beta)>=|A_{\pm}(\beta)>.}}

3. The  n-particle in-states and out-states\ \hrt\  are regarded
respectively as the spaces
\ ${\cal V}_n\otimes\cdots\otimes{\cal V}_1$\  and
\ ${\cal V}_1\otimes\cdots\otimes{\cal V}_n$\ .
Actions of the
generators\ $Q^{\pm},Q_{0}$\ are specified  by the coproduct:
 \eqn\ou{\Delta(H)=1\otimes H +H \otimes 1,}
\eqn\khjfg{\Delta(X^{\pm})=X^{\pm}\otimes 1-1\otimes X^{\pm}.}
It is easy to check that this  prescription  introduces a structure of
\ $sl(2)$\ representation
on the space\ $\pi_A$.

Precisely speaking, the unusual choice of the sign
in the coproduct\ \khjfg\  implies  that we are considering the representation
of the quantum algebra\ $ U_{-1}(sl(2))$. Let us discuss the reason of our
choice. First of all, note  that the   in- and out-spaces
\ ${\cal V}_2\otimes{\cal V}_1,\  {\cal V}_1\otimes{\cal V}_2$\
are isomorphic, and that the two particles S-matrix
defines the operator
\eqn\opiu{\check S:{\cal V}_2\otimes
{\cal V}_1\rightarrow {\cal V}_1\otimes{\cal V}_2.}
Using the explicit form of the S-matrix\ \jvhbf\  , one can represent
this operator in the form:
\eqn\rjsdh{\check S=S(\beta)\left(P_1+\frac{i \pi+\beta}{i \pi-\beta} P_0
\right),}
where\ $P_1$\ and\  $P_0$\ are  projectors on  three and one dimensional
\ $U_{-1}(sl(2))$-irreducible components in the tensor product
\ ${\cal V}_2\otimes{\cal V}_1$.
As it follows from \ \rjsdh\ ,
$\check S$\ commutes with
charges\ $Q^{\pm},Q^{0}$\ \vb\  and they are integrals of motion.

Thus the Quantum Field Theory
corresponding to FST\
\jvhbf\  contains  conserved currents\ $J_{\mu}^{m},
\{m=0,\pm;\mu=x,t\}$\  and
\eqn\mnj{\eqalign{&Q^0=\int_{-\infty}^{+\infty}d x J^0_t(x,t),\cr
&Q^{\pm}=\int_{-\infty}^{+\infty}d x J^{\pm}_t(x,t).}}
One can expect that they are  local operators in the theory.
It means that  commutators
$$\lbrack J_{\mu}^m(x),J_{\mu}^m(y)\rbrack$$
vanish on the spacelike
Minkowski interval \ $(x^{\mu}-y^{\mu})^2<0$\ . Moreover,
from the formulas for the coproduct\ \ou, \khjfg\  we conjecture that
the current\ $J_{\mu}^{0}$\ is a local operator and currents $J_{\mu}^{\pm}$\
are semilocal ones
with respect the "elementary" field $\mit\Psi_a(x)$.
The "elementary" field $\mit\Psi_a(x)$\ means any field with nonzero
matrix elements between vacuum and one particle states.

Let me recall here  the definition of a mutual locality index.
Consider  the  operator product
$$A(x^{\mu})B(y^{\mu}).$$
As a function of the  variable \ $x$\ it can have nontrivial
monodromic properties.  Suppose
\eqn\hfg{{\cal A}_C\big[A(x^{\mu})B(y^{\mu})\big]=\exp(2\pi i\omega(A,B))
A(x^{\mu})B(y^{\mu}).}
Here the symbol \ ${\cal A}_C$\ means the analytical continuation
along a counterclockwise contour \ $C$\  around the point \ $y^{\mu}$\ .
Then the number\ $\omega(A,B)$\ is  the  mutual
locality index  for the fields \ $A$\ and\ $B$\ .
In the present case
\eqn\hdgfs{\omega(J_{\mu}^0,\mit\Psi )=0 \ ,\ \ \
\omega(J_{\mu}^{\pm},\mit\Psi )=\frac{1}{2}.}

It is well known that   FST\ \jvhbf\
corresponds to \ $ SU(2)$\ TM,
which can be defined by the Lagrangian\  \pea ,\ \bank ,\ \hal
\eqn\ljoi{L= \int_{-\infty}^{+\infty}d x
(i\ \bar\psi\gamma^{\mu}\partial_{\mu}
\psi-g J_{\mu}^m J_m^{\mu}).}
The fields $\psi=\{\psi^{a}_{i}\}$\ are Dirac spinors with
the spinor index \ $\{i=1,2\}$\ and isotopic index \
$\{a=1,2\}$\ and the currents\ $J_{\mu}^{m}(x)$\ are equal to
\eqn\opiusq{\eqalign{&J^0_{\mu}=
\frac{1}{\sqrt2}\bar\psi\gamma_{\mu}\sigma^3\psi,\cr
&J^{\pm}_{\mu}=\frac{1}{2}\bar\psi\gamma_{\mu}(\sigma^1\pm i\sigma^2)\psi.}}
Here the Pauli matrices \ $\sigma^{k},\{ k=1,2,3\} $\ act on isotopic indices
of  spinors.

On the classical level the theory\ \ljoi\  is conformally
invariant. However,
its quantum spectrum contains a free massless boson
and two massive kinks\ $A_a$\ forming the fundamental multiplet
of the isotopic group\ $ SU(2)$.
The two-particles S-matrix for the  kinks
is determined  by \ \jvhbf .

Another non trivial example of FST is  connected with    SGM\  \col ,\ \ztt ,
whose
Lagrangian is given by
\eqn\sine{L=
\int_{-\infty}^{+\infty}d x
\big[\frac{1}{2}(\partial_{\mu}\varphi)^2+\frac{m^2_0}{b^2}\cos(b\varphi)
\big].}
If the parameter
\eqn\mbnn{\xi=\frac{b^2}{8\pi-b^2}}
is more than one,
the quantum spectrum of  SGM contains only solitons  $A_a$
possessing an internal degree of freedom $ a=\pm1 $\ (soliton-antisoliton).
Their two-particle \ $S$-matrix
was found in the pioneering work\ \z\ . It
reads explicitly
\eqn\jvhbfsdq{\eqalign{&S_{++}^{++}(\beta)=S_{--}^{--}(\beta)=S(\beta),\cr
&S_{+-}^{+-}(\beta)=S^{-+}_{-+} (\beta)=S(\beta)\frac{\sinh\frac{ \beta}{\xi}
}{\sinh\frac{i \pi-\beta}{\xi}},\cr
&S_{+-}^{-+}(\beta)=S^{+-}_{-+}(\beta)=S(\beta)\frac{\sinh\frac{ i\pi}{\xi}}
{\sinh\frac{i  \pi-\beta}{\xi}}.}}
Here
\eqn\bvvc{S(\beta)=-\frac{\Gamma(\frac{1}{\xi})\Gamma(1+\frac{i\beta}{\pi\xi})}
{\Gamma(\frac{1}{\xi}+\frac{i\beta}{\pi\xi})}
\prod_{p=1}^\infty\frac{R_p(\beta)R_p(i\pi-\beta)}{R_p(0)R_p(i\pi)},}
$$R_p(\beta)=\frac{\Gamma(\frac{2 p}{\xi}+\frac{i\beta}{\pi\xi})
\Gamma(1+\frac{2 p}{\xi}+\frac{i\beta}{\pi\xi})}
{\Gamma(\frac{2 p+1}{\xi}+\frac{i\beta}{\pi\xi})
\Gamma(1+\frac{2 p-1}{\xi}+\frac{i\beta}{\pi\xi})}.$$
The charge conjugation matrix \ $\C_{ab}$\
for  this FST  has the  form\ \char.

The Hilbert space of
asymptotic states for  SGM \  possesses a symmetry
described by the quantum algebra\ $U_q(sl(2))$\
\ref\lecl{A. LeClair, Phys. Lett. B230 (1989) 103. },
\ref\resm{N.Yu. Reshetikhin and F.A. Smirnov, Commun. Math. Phys. 131
(1990) 157.},
\lec\
with
\eqn\qwe{q=\exp(i\pi\frac{1+\xi}{\xi}).}
To fix the notation let me recall that the algebra
\ $U_q(sl(2))$\ is  generated by operators\ $E^{\pm}$,\  $H$\
with the  commutation relations\
\ref\rks{N.Yu. Reshetikhin, P.P. Kulish and E.K. Sklyanin, Lett.Math.
Phys. 5 (1981) 393.},
\ref\drin{V.G. Drinfel'd, Quantum Groups, Proc. ICM-86 (Berkeley,  CA),
1 (1987) 798.} :
\eqn\tyrrte{\eqalign{&[H,E^{\pm}]=\pm \sqrt2 E^{\pm},\cr
&[E^{+},E^{-}]=\frac{q^{\sqrt2  H}-q^{-\sqrt2  H}}
{q-q^{-1}}}}
and  coproduct
\eqn\tres{\eqalign{&\Delta(H)=1\otimes H +H \otimes 1, \cr
&\Delta(E_{\pm})=q^{-\frac{\sqrt2 }{2}H}\otimes E_{\pm}+
E_{\pm}\otimes q^{\frac{\sqrt2 }{2}H}.}}

The standard realization of \ $U_q(sl(2))$-invariance in  SGM implies
that we have to identify the vectors
$$\hat A_a(\beta)|vac>,$$
where
\eqn\mbngh{\hat A_a(\beta)=\exp(a \frac{\beta}{2 \xi}) A_a(\beta), }
with the basis of the fundamental representation of   \ $U_q(sl(2))$.
The commutation relations for the operators\  \ \mbngh\ are  defined
by the matrix
\eqn\lgki{\hat S^{cd}_{ab}(\beta)=
\exp\big(\frac{(a-c)\beta}{2 \xi}\big)
S^{cd}_{ab}(\beta).}
One can prove that the operator\ $\check S$\ \opiu\    corresponding
to  the matrix\ \lgki\
commutes with the action of the charges
\eqn\bdyhf{\eqalign{&Q^{\pm}=\pi_A[E^{\pm}q^{\pm\frac{\sqrt2 }{2}H}],\cr
&Q^0=\pi_A[H]\ .}}

It is necessary to point out that there is
an essential difference between \ $ SU(2)$\ TM and  SGM .
The currents \ $J_{\mu}^{\pm} $\ corresponding to the  conserved
charges\ $Q^{\pm}$\
are not local operators in  SGM.
At the same time the current  \ $J_{\mu}^{0} $\  generating the
\ $U(1)$-charge \ $Q^0$\
is a local one. It is also  mutually local with respect
the "elementary" field.

The structures of local commutative IM
\ $I_s$\ are identical
for  \ $SU(2)$\ TM and SGM.
The  function\ $s(\alpha)$\ \gdfsd\  was
found in the work \
\ref\sk{E.K. Sklyanin and L.D. Faddeev, Sov. Phys. Dokl, 23 (1978) 902.}
\eqn\jdhfg{s(\alpha)=\coth\frac{\alpha}{2}\ .}
As it  follows from this formula, the spins of the
local commutative IM  are
\ $s=1$\ (mod 2).
Note that the first ones
$$I_1-I_{-1}\ ,\ \  I_1+I_{-1}$$
are the momentum and energy charges.

\subsec{Form-factors}

In many cases the FST  data are sufficient
to reconstruct  matrix elements
\eqn\form{F^{b_1...b_m}_{a_1...a_n}
(\beta'_1,...\beta'_m|\beta_1,...\beta_n)=
_{out}<A^{b_m}(\beta'_m)...A^{b_1}(\beta'_1)|O(0)|A_{a_1}(\beta_1)...A_{a_
n}(\beta_n)>_{in},}
of an hermitian local operator $O(x)$  between asymptotic
states\ \kar, \smirn\ .
It is convenient to introduce the following functions, called form-factors
\eqn\firm{F_{a_1...a_n}(\beta_1,...\beta_n)=
<vac|O(0)|A_{a_1}(\beta_1),...A_{a_n}(\beta_n)>_{in},}
which are  matrix elements of an operator\
 $ O(x) $\  at the origin  between
an
n-particle in-state and the vacuum state .
Crossing symmetry  implies that
a general matrix element  \form\
is obtained by an analytical continuation of \firm  , and equals \ \kar
\eqn\furm{F^{b_1...b_m}_{a_1...a_n}
(\beta'_1,...\beta'_m|\beta_1,...\beta_n)=
\C^{b_1c_1}...\C^{b_mc_m}
F_{a_1...a_nc_1...c_m}(\beta_1,...\beta_n,\beta'_1+i\pi,...\beta'_m+i\pi).}
Thus, to describe the local operator one should present a set of tensor valued
functions \firm. Their reconstruction is based on the following system of
axioms\ \smirn, \yur;

{\bf Axioms}

1. Function $F_{a_1...a_n}$$(\beta_1,...\beta_n)$ is analytic
in variables  \  $\beta_{ij}=\beta_i-\beta_j$\ inside the strip
\ $0<\Im m \beta<2\pi$\
except for simple poles. It becomes
the physical matrix elements\ \firm\
when all\ $\beta_i$\
are real and ordered as follows
$$\beta_1<\beta_2<...<\beta_n\ .$$

2. Relativistic invariance demands that form-factors satisfy
the equation
\eqn\si{F_{a_1...a_n}(\beta_1+\theta,\beta_2+\theta,...\beta_n+\theta)=
\exp(\theta s(O))F_{a_1...a_n}(\beta_1,...\beta_n),}
where\ $s(O)$\ is the spin of the local operator\ $O(x)$\ .

3. Form-factors should satisfy the  symmetry property (Watson's theorem)
\eqn\mars{\eqalign{&F_{a_1...a_{j+1}a_j...a_n}
(\beta_1,...\beta_{j+1},\beta_j,...\beta_n)=\cr
&=S^{c_jc_{j+1}}_{a_ja_{j+1}}(\beta_j-\beta_{j+1})
F_{a_1...c_jc_{j+1},...a_n}(\beta_1,...\beta_j,\beta_{j+1},...\beta_n)}}

4. Form-factors satisfy the  equation
\eqn\sdv{F_{a_1...a_n}(\beta_1,...\beta_{n-1},\beta_n+2\pi i)=
\exp(2\pi i \omega (O,\mit\Psi))
F_{a_na_1...a_{n-1}}(\beta_n,\beta_1,...\beta_{n-1}),}
where the shift by $\ 2\pi i$\  is understood as an analytical continuation
and the number\ $\omega(O,\mit\Psi)$ \ means the mutual locality index\
\hfg\ for the
operator \ $O(x)$\ and the "elementary" field\ $ \mit\Psi$.

5. Form-factors\ $F_{a_1...a_n}(\beta_1,...\beta_n)$\
being
considered as a function of \ $\beta_n$\
have simple poles at the points
\ $\beta_n=\beta_j+i\pi$\  with the following
residues
\eqn\bigger{\eqalign{
&i\  F_{a_1...a_n}(\beta_1,...\beta_n)=\C_{a_na'_j}
\frac{F_{a'_1...\hat a'_j...a'_{n-1}}
(\beta_1,...\hat \beta_j,...\beta_n)}{\beta_n-\beta_j-\pi i}\times\cr
&\big[\delta^{a'_1}_{a_1}...
\delta^{a'_{j-1}}_{a_j}S^{a'_{n-1}a'_j}_{a_{n-1}c_1}(\beta_{n-1}-\beta_j)
S^{a'_{n-2}c_1}_{a_{n-2}c_2}(\beta_{n-2}-\beta_j)\times\cr
&...S^{a'_{j+1}c_{n-j-2}}_{a_{j+1}a_j}(\beta_{j+1}-\beta_j)
-\exp(2 \pi i \omega(O,\mit \Psi))
S^{a'_ja'_1}_{c_1a_1}(\beta_j-\beta_1)\times\cr
&...S^{c_{j-3}a'_{j-2}}_{c_{j-2}a_{j-2}}(\beta_j-\beta_{j-2})
S^{c_{j-2}a'_{j-1}}_{a_ja_{j-1}}(\beta_j-\beta_{j-1})
\delta^{a'_{j+1}}_{a_{j+1}}...\delta^{a'_{n-1}}_{a_{n-1}}\big]+
...\ \ .}}
In the absence of bound states these poles are the only singularities of
\ $F_{a_1...a_n}(\beta_1,...\beta_n)$\
in the  strip\ $0<\Im m\beta_j<2\pi$\
for real \ $\beta_1,..\hat \beta_{j},...\beta_{j_n}$\ .

It was shown \ \ki,\ \smirn\  that the operators \ $O(x)$\  defined by
the matrix elements\ \firm\ satisfy  locality
relations
provided
the form-factors satisfy\ 1-5\ .

Once the form factors of the theory are known,  correlation
functions of local operators can be written as an infinite series over
multi-particle intermediate states. For instance, the two point
function of an operator\ $O(x)$\ for the spacelike Minkowski
interval \ $(x-y)^2=-r^2$\
is given by
\eqn\uyt{\eqalign{&<O(x)O(y)>=\cr
&\sum^{\infty}_{n=0}\int\frac{d \beta_1...d \beta_n}
{n! (2 \pi)^n}
F_{a_1...a_n}(\beta_1,...\beta_n)
F^{a_n...a_1}(\beta_n,...\beta_1)\exp(-r m \sum^n_{k=1}\cosh\beta_k).}}
All the integrals are non singular and convergent. The series is expected
to be convergent as well. Similar expressions can be derived for
multi-point correlators.

\newsec{ Form-factors reconstruction}

The system of form-factors axioms (1-5) is a
complicated Riemann-Hilbert problem. Its solution can be based
on the following  idea.

Let us assume that we have the representation\ $\pi_Z$\
of the formal Zamolodchikov-Faddeev
algebra, which satisfies the  requirements;

1. In the space\ $\pi_Z$\
an action of the operators \ $Z_a(\beta)=\pi_Z[V_a(\beta)]$ is defined
\eqn\uytras{Z_{a}(\beta_1) Z_{b}(\beta_2)=
S^{cd}_{ab}(\beta_1-\beta_2)
 Z_{d}(\beta_2) Z_{c}(\beta_1).}

2. The singular part of the operator product
$$Z_a(\beta_2)Z_b(\beta_1),$$
being considered as a function of the complex  variable
\ $\beta_2$\ for real \ $\beta_1$\
in the upper
half plane $\Im m \beta_2\geq 0$\ , contains only  simple pole with
the residue
\eqn\al{  i\ Z_a(\beta_2)Z_b(\beta_1)=
\frac{\C_{ab}}{\beta_2-\beta_1-\pi i}+... .}
It means that there is only one singularity \al\   depending on
the real parameter \ $\beta_1$\ for  a general matrix  element
$$<u|Z_a(\beta_2)Z_b(\beta_1)|v>,\ |u>,|v>\in \pi_Z.$$
Of course, this matrix element may also have other singularities
for  \ $\Im m \beta_2\geq0$\ , but their  positions  are defined
by the  vectors   \ $|u>$\ and\ $|v>$\  only.

3. There is a unique G-invariant
\foot{ recall  that G is a finite dimensional symmetry group of the model  }
 vector\ $|0>$\  (vacuum state) in the space
\ $\pi_Z$\ such that
the two point function
\eqn\gdgf{G_{ab}(\beta_1,\beta_2)=
<0|Z_a(\beta_2) Z_b(\beta_1)|0>,}
satisfies the following constraints:

\ \ a. It depends only on the difference\ $\beta=\beta_1-\beta_2$\

\ \ b. As a function of the complex  variable\ $\beta$\  ,
it is  analytic
in the  lower

\ \ half plane\ $\Im m \beta\leq 0$\  except for one simple pole\ \al\ .

\ \ c. It is a bounded  function for \  $ \beta\to\infty , \Im m\beta\leq 0$
\foot{on  this requirement  see the comment at the end of the next section}
\eqn\ogr{G_{ab}(\beta)=O(1) ,\ \
\beta\rightarrow\infty\ (\Im m\beta\leq0).}

  The representation\ $\pi_Z$\  will be completely specified
if  all vacuum matrix elements
\eqn\tersf{G_{a_1...a_n}(\beta_1,...\beta_n)=
<0|Z_{a_n}(\beta_n)... Z_{a_1}(\beta_1)|0>}
are defined. The problem of their reconstruction
is  close to the
Riemann-Hilbert problem (1-5).

To clarify this connection
let us suppose that
the following additional structures are present  in the
space\ $\pi_Z$\ ;

1. The operator \ $K$\ , which obeys the commutation relation
\eqn\bcvcl{Z_a(\beta+\theta)=\exp(-\theta K)Z_a(\beta)\exp(\theta K);}

2. The map
\eqn\mvnbb{ O\to  \Lambda(O)\in {\rm End}[\pi_Z]}
from the space of local operators  of the  model under investigation
to the endomorphizm algebra of \ $\pi_Z$\ , which satisfies the
conditions:
\eqn\lor{\eqalign{&\Lambda(O)Z_a(\beta)=
\exp(2\pi i \omega(O,\mit\Psi))Z_a(\beta)\Lambda(O),\cr
&\exp(\theta K)\Lambda(O)\exp(-\theta K)=\exp(\theta s(O))\Lambda(O),}}
where the number\ $\omega(O,\mit\Psi)$\   coincides with the
mutual locality index\ \hfg\
of the "elementary" field\ $\mit\Psi(x)$\  and the
local operator\ $O(x)$\ with the
spin \ $s(O)$\ .

Using the cyclic properties of a matrix trace and the
relations \ \uytras,\ \al,\ \lor\ it is easy to verify that the function
\eqn\tr{F_{a_1...a_n}(\beta_1,...\beta_n)=\Tr_{\pi_Z}\big[\exp(2\pi i K)
\Lambda(O)Z_{a_n}(\beta_n)...Z_{a_1}(\beta_1)\big]}
formally satisfy the
axioms\ \si -\bigger.
Certainly this  observation
is not the rigorous   method to solve
the Riemann-Hilbert problem \ (1-5),
since  we have no a  proof that the
trace\ \tr\ exists and satisfies  axiom 1. Nevertheless, it can be
substantiated in some cases. In this paper it will be demonstrated for
\ $SU(2)$\ TM and SGM.

At the conclusion of this section let us briefly argue  a
physical interpretation of the space \ $\pi_Z$\  \zamolodc.
First of all, note that the space \ $\pi_A$\
is associated with an infinite line\ $t=const$.
The formula\ \tr\  implies that \ $\pi_A$\  is
obtained by  "gluing"
two copies of   \ $\pi_Z$.
So we can  associate the space    \ $\pi_Z$\
with half infinite line.
In  other words
one should consider  \ $\pi_Z$\ as  a space of   angular
quantization of a massive integrable model
\ref\lecla{A. LeClair, Spectrum Generating Affine Lie Algebras
and Correlation Functions in
Massive Field Theory, Cornell preprint (1993)
CLNS 93/1220.}.

\newsec{Two point vacuum averages }

To describe the representation \ $\pi_Z$\ of the  Zamolodchickov-
Faddeev algebra for  \ $SU(2)$\ TM and SGM we have to reconstruct
all vacuum averages
$$G_{a_1...a_n}(\beta_1,...\beta_n)=
<0|Z_{a_n}(\beta_n)... Z_{a_1}(\beta_1)|0>.$$
Due to the unbroken \ $U(1)$-symmetry present in the
models under investigation,
they are non trivial only for even n.
In this section
the two point functions
\eqn\gdgf{G_{ab}(\beta_1-\beta_2)=
<0| Z_a(\beta_2) Z_b(\beta_1)|0>}
will be found.

First, let us consider  the case of
\ $SU(2)$\ TM.
The commutation relation\ \uytras\  means that \ \gdgf\
obeys the functional
equation:
\eqn\reewdfs{G_{ab}(-\beta)=S_{ab}^{cd}(\beta) G_{dc}(\beta).}
It is convenient to introduce the  function
\eqn\yt{g(\beta)=k^{\frac{1}{2}} \frac{\Gamma(\frac{1}{2}+\frac{i\beta}{2\pi})}
{\Gamma(\frac{i\beta}{2\pi})},}
where \ $k$\ is an arbitrary number.
It satisfies the  following:

a. It is an analytical function
the lower half plan\ $\Im m \beta\leq 0$;

b. If\ $\beta\rightarrow\infty\ (\Im m \beta\leq 0)  $\ , then
\eqn\bvhf{g(\beta)=\kappa^{\frac{1}{2}}(i \frac{\beta}{2\pi} )^{\frac{1}{2}}
(1+O(\frac{1}{\beta})).}

c. The function\ $g(\beta)$\ obeys the functional equation:
\eqn\gdfu{S(\beta)=\frac{g(-\beta)}{g(\beta)}.}

As it follows from \ \rjsdh\ , the general solution of the
equation\  \reewdfs\ is  given by:
\eqn\rews{\eqalign{&G_{ab}(\beta)=\frac{g(\beta)}{g(-i\pi)}
\Biggr(\left[\matrix{\frac{1}{2}&\frac{1}{2}&0\cr
\frac{a}{2}&\frac{b}{2}&0}\right]_{-1}\frac{A(\beta)}{i\pi+\beta}+
\left[\matrix{\frac{1}{2}&\frac{1}{2}&1\cr
\frac{a}{2}&\frac{b}{2}&\frac{a+b}{2}}\right]_{-1} B(\beta)\Biggr).}}
Here \ $A(\beta)$\ and \ $B(\beta)$\ are  arbitrary even functions,
and  the symbol
$$\left[\matrix{j_1&j_2&j \cr m_1&m_2&m}\right]_q$$
denotes  Clebsch-Gordan coefficient
for the  quantum universal enveloping algebra \
$U_q(sl(2))$
\ref\kirre{A.N. Kirillov and N.Yu. Reshetikhin,
ed. V.G. Kac, World Scientific, Singapore (1989).}
\foot{in this work the normalization of
Clebsh-Gordan coefficients is chosen as in   Ref.\kirre\  }
, in  particular
\eqn\ter{\eqalign{&\left[\matrix{\frac{1}{2}&\frac{1}{2}&0\cr
\frac{a}{2}&\frac{b}{2}&0}\right]_{-1}=\frac{\delta_{a+b,0}}{\sqrt2},\cr
&\left[\matrix{\frac{1}{2}&\frac{1}{2}&1\cr
\frac{a}{2}&\frac{b}{2}&\pm1}\right]_{-1}=\delta_{a+b,\pm1},\cr
&\left[\matrix{\frac{1}{2}&\frac{1}{2}&1\cr
\frac{a}{2}&\frac{b}{2}&0}\right]_{-1}=
(-1)^{\frac{1-a}{2}}\frac{\delta_{a+b,0}}{\sqrt2}.}}

Due to the \ $U_{-1}(sl(2))$-invariance of the  vacuum state\ $|0>$ ,
the function \ $B(\beta)$\ \rews\  must be zero
\eqn\gfdxl{B(\beta)\equiv0.}
At the same time, the analyticity condition implies  that the even
function\ $A(\beta)$\ is   analytic  in the  whole complex
plane and
$$A(-i\pi)=i \sqrt2.$$
Moreover, the boundary condition\ \ogr\  provides the unique reconstruction
of this function:
\eqn\lop{A(\beta)\equiv i \sqrt2.}

We should note at this point
that the boundary condition\ \ogr\   for the two point function
\ $G_{ab}(\beta)$\ is closely  connected  with the
unbroken symmetry condition. Indeed if the function
\ $B(\beta)$\ \rews\   is non zero, we could not impose\ \ogr  .

In this way,  the two point function for  \ SU(2)\ TM is  given by
\eqn\mnvf{G_{ab}(\beta)=\frac{g(\beta)}{g(-i\pi)}
\frac{ i\delta_{a+b,0}}{i\pi+\beta}.}

One can find the two point function for  SGM  in a
similar fashion. I have
to make the following comment only;
To apply  the boundary condition\  \ogr\
one  have to use
the basis of the Zamolodchikov-Faddeev operators
which conforms to the G-invariance of the  theory.
So, in the case of   SGM we have to  consider the following simple
redefinition of the Zamolodchikov-Faddeev operators (compare with \ \mbngh) :
\eqn\cxes{\hat Z_{a}(\beta)=\exp( \frac{a \beta}{2 \xi}) Z_{a}(\beta).}
Then the function
\eqn\bdfsr{\hat G_{ab}(\beta)=\exp\big(- \frac{a\beta}{2 \xi}\big)
G_{ab}(\beta)}
will satisfy \  \ogr\ .
Here is its explicit form
\eqn\bvh{\hat G_{ab}(\beta)=
(-1)^{\frac{a+1}{2}} q^{\frac{a}{2}} \delta_{a+b,0}
\frac{ g(\beta)}{g(-i \pi)}
\frac{\Gamma(1+\frac{2}{\xi})\Gamma(-\frac{1}{\xi}+\frac{i\beta}{\pi\xi})}
{i\pi \xi\Gamma(1+\frac{1}{\xi}+\frac{i\beta}{\pi\xi})},}
where
\eqn\nbv{g(\beta)=\left[\frac{\kappa}
{\Gamma(\frac{1}{\xi})}\right]^{\frac{1}{2}}
\frac{\Gamma(\frac{1}{\xi}+\frac{i\beta}{\pi\xi})}
{\Gamma(\frac{i\beta}{\pi\xi})}
\prod_{p=1}^{+\infty}\frac{\left[R_p(i\pi)R_p(0)\right]
^{\frac{1}{2}}}{R_p(\beta)},}
The functions\ $R_p(\beta)$\ are defined by the equation\ \bvvc \
and \ $\kappa$\  is an arbitrary constant.

\newsec{Bosonization technique for massive integrable models}

In the following
I shall have to introduce a lot of new functions and constants.
Their explicit expressions are  complicated  enough and essentially
depend on a model. In order to avoid  bulky
formulas I will try to  use  unique symbolic notations for
these objects and
point out their universal properties.
One can find the explicit forms of  the introduced functions and constants
in the  Appendices 2,3.

\subsec{Free boson field}

Consider the formal operator-valued
function \ $ \phi (\beta) $\
which obeys the  commutation
relation for real\ $\beta$\ :
\eqn\rimaq{[\phi(\beta_1),\phi(\beta_2)]=\ln S(\beta_2-\beta_1) .}
In order to construct the representation of \ \rimaq\
it is necessary to specify the two point
correlation function
\eqn\tdfsr{<0|\phi(\beta_1)\phi(\beta_2)|0>=-\ln g(\beta_2-\beta_1).}
The compatibility of \  \tdfsr\  and
\ \rimaq\ means that the functions \ $g(\beta),\ S(\beta)$\ are
connected by the relation:
\eqn\gdfsre{S(\beta)=\frac{g(-\beta)}{g(\beta)}.}
We will demand that \ $g(\beta)$\ satisfies also
the proper analyticity and boundary conditions,
which are the natural generalization of   properties
of the  free boson Green function in CFT.

a. It is an analytical function  without
zeroes and poles in the lower half plane\ $\Im m\beta\leq 0$ except
simple zero at \ $\beta=0$.

b. If\ $\beta\to\infty\ (\Im m\beta\leq 0), $\ then
$$\partial_{\beta} \ln g(\beta)=O(\frac{1}{\beta}).$$
The equation \ \gdfsre\   supplemented with the conditions (a-b) is
the simplest Riemann-Hilbert problem.
It has the unique solution\  \yt .

Now, let us introduce   the field
\eqn\rasa{\bar\phi(\beta)=
\phi(\beta+i\frac{\pi}{2})+\phi(\beta-i\frac{\pi}{2}).}
It should be considered as a more fundamental
object than the field\ $\phi(\beta)$\ , since it
has more simple and universal properties.
For example,
using the unitarity  and crossing symmetry equations
\eqn\tsre{\eqalign{&S(\beta)S(-\beta)=1,\cr
&S(i\pi-\beta)=\frac{\beta}{i \pi-\beta} S(\beta)}}
one  can get the commutation relations for this field
\eqn\rim{[\bar \phi(\beta_1),\bar \phi(\beta_2)]=
\ln\frac{\beta_2-\beta_1-i\pi}{\beta_2-\beta_1+i\pi} ,}
\eqn\hys{[\bar \phi (\beta_1),\phi (\beta_2)]=
\ln\frac{\frac{i\pi}{2}-\beta_2+\beta_1}{\frac{i\pi}{2}+\beta_2-\beta_1}.}
The two point functions
\eqn\hfgxbn{<0|\bar \phi(\beta_1)\phi(\beta_2)|0>
=\ln w(\beta_2-\beta_1),}
\eqn\mnvb{<0|\bar \phi(\beta_1)\bar\phi(\beta_2)|0>
=-\ln \bar g(\beta_2-\beta_1).}
also have simple forms:
\eqn\hdbc{\eqalign{&w(\beta)=k^{-1} \frac {2 \pi}{i (\beta+i\frac{\pi}{2})}
,\cr
&\bar g(\beta)=-k^2 \frac{\beta (\beta+i \pi)}{4 \pi^2}.}}
Notice  that the functions\ $g(\beta),\ \bar g(\beta),\
w(\beta)$\ are connected as follows :
\eqn\gffd{\eqalign
{&w(\beta)=
\left[g(\beta+i\frac{\pi}{2})g(\beta-i\frac{\pi}{2})
\right]^{-1},\cr
&\bar g(\beta)=
\left[
w(\beta+i\frac{\pi}{2})w(\beta-i\frac{\pi}{2})
\right]^{-1}.}}

\subsec{ Elementary vertex operators}

Consider the elementary vertex operators
\eqn\vert{\eqalign{&V(\beta)=\exp(i\phi (\beta))=(g(0))^{\frac{1}{2}}
 :\exp(i \phi(\beta):\
,\cr
&\bar V(\beta)=\exp (-i\bar \phi (\beta))=
(\bar g(0))^{\frac{1}{2}}: \exp (-i\bar \phi(\beta)):\ ,}}
here the dots implies the normal ordering of  exponents.
\foot{ since the two point functions are c-numbers
Wick theorem can be applied}
The functions\ $g(\beta)$\
and \ $\bar g(\beta)$\ have simple zeroes when\ $\beta=0,$\
hence  the vertex operators \ \vert\  should be regularized.
We shall do it in a similar fashion as in 2D CFT.
It means that regularized values of the  functions  \ $g(\beta)$\ and
\ $\bar g(\beta)$\
for \ $\beta=0$\ are  the limits

\eqn\fers{\eqalign{& g_{ reg}(0)=
lim_{\scriptstyle \beta\to 0 }\frac{g(\beta)}{\beta}\equiv\rho^2,\cr
&\bar g(0)_{reg}=lim_{\scriptstyle \beta\to 0 }
\frac{\bar g(\beta)}{\beta}\equiv\bar
\rho^2.}}

The correlations functions\ \tdfsr,\ \hfgxbn,\ \mnvb\
define  expressions of the  operator products:
\eqn\reaf{\eqalign{&V(\beta_1)V(\beta_2)=\rho^2 g(\beta_2-\beta_1):V(\beta_1)
V(\beta_2):,\cr
&\bar V(\beta_1)V(\beta_2)=\rho\bar\rho
w(\beta_2-\beta_1):\bar V(\beta_1)
V(\beta_2):,\cr
&\bar V(\beta_1) \bar V(\beta_2)=\bar \rho^2 \bar g(\beta_2-\beta_1)
:\bar V(\beta_1)
\bar V(\beta_2):.}}

\subsec{"Screening charge"}

In the space of representation\ $\pi_Z$\  of the algebra\ \rimaq\
( it will be  accurately described later )
one can introduce the  operator \ ${\cal X}$\
\eqn\terfs{<u|{\cal X}|v>\equiv\eta^{-1}<u|
\int_{\scriptstyle C}
\frac{d\gamma}{2\pi}
\bar V(\gamma)|v>,}
where \ $ \eta$\  is an irrelevant constant.
The contour of integration\ $C$\ is specified  as follows;

First of all, we assume that
matrix elements
$$<u|\bar V(\gamma)|v>$$
are  meromorphic functions decreasing  infinitely faster
than \ $\gamma^{-1}$\  for all vectors\ $|u>,|v>\in\pi_Z$.
Then the contour\ $C$\  goes from\ $ \Re e\ \gamma=-\infty$\ to
\ $\Re e\ \gamma=+\infty$. It       lies
above all singularities whose positions   depend on the
vector\ $|v>$\ , but below singularities  depending on  \ $|u> $.

To illustrate this definition let us
calculate the following matrix elements of the operator\ ${\cal X}$
\eqn\sdwe{\eqalign{&J(\beta_1-\beta_2)
=<0|V(\beta_2){\cal X}V(\beta_1)|0>,\cr
&J_1(\beta_1-\beta_2)=<0|V(\beta_2)V(\beta_1){\cal X}|0>,\cr
&J_2(\beta_1-\beta_2)=<0|{\cal X} V(\beta_2)V(\beta_1)|0>.}}
{}From the formula \ \reaf\ we immediately obtain
the integral representation for these
matrix elements
\eqn\fsrd{\eqalign{&J(\beta_1-\beta_2)=\eta^{-1}\rho^2\bar\rho
g(\beta_1-\beta_2)\int_{\scriptstyle C}\frac{d\gamma}{2\pi}
w(\gamma-\beta_2)w(\beta_1-\gamma),\cr
&J_1(\beta_1-\beta_2)=\eta^{-1}\rho^2\bar\rho
g(\beta_1-\beta_2)\int_{\scriptstyle C_1}\frac{d\gamma}{2\pi}
w(\gamma-\beta_2)w(\gamma-\beta_1),\cr
&J_2(\beta_1-\beta_2)=\eta^{-1}\rho^2\bar\rho
g(\beta_1-\beta_2)\int_{\scriptstyle C_2}\frac{d\gamma}{2\pi}
w(\beta_2-\gamma)w(\beta_1-\gamma).}}
In the  case at hand
the contours of integration are specified by the prescription
(see formula\ \hdbc\ ):

The contour\  $C$\  lies above the  pole \ $\gamma=\beta_1+i\frac {\pi}{2},$\
but bellow $\ \gamma=\beta_2-i\frac {\pi}{2}$\ ;

The contour\ $ C_1$\
lies bellow the  poles \ $\gamma=\beta_{1,2}-i\frac {\pi}{2}$\ ;

The contour \ $C_2$\  lies above the poles\
$\gamma=\beta_{1,2}+i\frac {\pi}{2}.$

The functions\ \fsrd\ are  given by:
\eqn\tersxfs{\eqalign{&J(\beta)=\frac{ 2\pi\bar \rho }{  \eta k }
\frac{g(\beta)}{g(-i\pi)}\frac{1}{\beta+i\pi},\cr
&J_1(\beta)=J_2(\beta)=0.}}
It is convenient to chose the parameter \ $\eta$\ such  that \ $
\frac{2 \pi \bar \rho }{\eta k }=1$\ , so
\eqn\red{\eta=\sqrt{\frac{\pi}{i}}.}
Note that there is the  simple relation between the functions
\mnvf\ and\ \tersxfs\
\eqn\rwesd{G_{ab}(\beta)=i J(\beta)\delta_{ab}.}

\subsec{Free field representation  for the Zamolodchikov-Faddeev algebra
of  \ $ SU(2)$\
TM }

Define the operators\ $Z_a(\beta)$\ for
\ $SU(2)$\ TM by the formulas:
\eqn\ters{\eqalign{&Z_+(\beta)=V(\beta),\cr
&Z_-(\beta)=i({\cal X}V(\beta)+V(\beta)
{\cal X}).}}
The following propositions  describe  commutation relations
and analytical properties of \   $\ Z_a(\beta)$\ .

{\bf Proposition 1.}

The operators\ \ters\  satisfy the Zamolodchikov-Faddeev commutation
relations\ \uytras.

{\bf Proposition 2.}

The singular part of the operator product
$$ Z_a(\beta_2) Z_b(\beta_1)$$
considered as a function of the
complex variable \ $\beta_2$\ for real \ $\beta_1$\
in the upper half plane \ $\Im m \beta_2\geq 0$\ contains
only one simple pole with
residue equals  \ $ \frac{1}{i} \C_{ab}.\   \spadesuit $

We  have proved these propositions
for  vacuum matrix elements in the last subsection.
The main steps of the  general proof  can be found in  Appendix 1.

The formulas\ \ters\
define
the integral representation for
n-point correlation functions\ \tersf\  for  \ $SU(2)$\ TM.
They  have simple group   theoretic
meaning.
Indeed, the operators\ $Z_a(\beta)$\
can be regarded  as the basis
of the fundamental representation of \ $U_{-1}(sl(2))$\ \tte .
If we
identify  \ ${\cal X}$\ with the \ $\pi_Z[X^{-}]$\ ,
it is easy to recognize the coproduct\ \khjfg\  in
the definition\ \ters . In the next section
the generators\ $\pi_Z[X^{+}]$\ and \ $\pi_Z[H]$\
will be constructed.

\newsec{Fock realization of the representation\ $\pi_Z$ }

In this section it will be  shown that the
space\ $\pi_Z$ can be represented in the form:
\eqn\lgjkh{\pi_Z=lim_{\scriptstyle \epsilon\to 0 }\pi^{\epsilon}_{Z}.}
Here the space \ $\pi_Z^{\epsilon}$\ admits a decomposition
into a direct sum of Fock modules and  \ $\epsilon$\
is a  parameter of the ultraviolet regularization.
In the case of \ $SU(2)$\  TM the
below construction is equivalent to the
Frenkel-Jing
bosonization of the quantum
affine algebra\ $U_q\widehat{( sl(2)})$\
 \fren,\ \japthree.
\subsec{Oscillator decomposition}
Trying to get an oscillator decomposition
for the field\ $\phi(\beta)$\ \rimaq\  , we run into the well known
problem.  The function
\ $\ln S(\beta)$\ is not tend to zero when\ $\beta\to\pm\infty$\ .
So, the commutation relation\ \rimaq\ does
not compatible with the  decreasing boundary condition for  the
field \ $\phi(\beta)$\  and
it
cannot be decomposed in to a  Fourier integral.

A  possible way out is to  consider the field \  $ \phi_{\epsilon}(\beta)$\
which is
defined on the finite interval
\eqn\lgjh{-\frac{\pi}{\epsilon}\leq\beta\leq\frac{\pi}{\epsilon}}
and satisfies the commutation relations:
\eqn\rimaqas{[\phi_{\epsilon}(\beta_1),\phi_{\epsilon}(\beta_2)]=
\ln S_{\epsilon}(\beta_2-\beta_1) .}
The function \ $S_{\epsilon}(\beta)$\ must tend to \ $S(\beta)$\
when\ $\epsilon\to 0$\ for finite \ $\beta$\   and
\eqn\kgjh{S_{\epsilon}(-\frac{\pi}{\epsilon})=S(-\infty)\equiv \exp(i\pi s).}
Then the field \ $ \phi_{\epsilon}(\beta)$\
admits the  following decomposition
\eqn\ytfh{\phi_{\epsilon}(\beta)=
s^{\frac{1}{2}}
(Q-
\epsilon\beta P)+
\phi_{\epsilon}^{osc}(\beta),}
where the field  \ $ \phi_{\epsilon}^{osc}(\beta)$\ is  periodic in the
interval\ \lgjh\  .
Zero modes \ $P,\ Q$\  obey the canonical
commutation relation
\eqn\tred{[P,Q]=\frac{1}{i}}
and commute with the oscillator part   \ $\phi_{\epsilon}^{osc}(\beta)$\ .
One should consider the operator\ $\phi(\beta)$\
as a properly
regularized limit of \  $ \phi_{\epsilon}(\beta)$.

Now let us apply  this simple idea to  our problem.
In the present  case the regularization\ \lgjh\  has a clear
physical meaning. Indeed we regard  the parameter \ $\beta$\
as  rapidity of physical excitations in massive models,
so  \lgjh\  is the ultraviolet
cut-off.

Introduce
the following expansion :
\eqn\teresa{\phi_{\epsilon}(\beta)=\sqrt{s}
(Q-\epsilon\beta P)+
\sum_{m\ne 0}\frac{a_m}{i\sinh( \pi m\epsilon)}\exp(i m \epsilon \beta).}
I wish to emphasize that this representation for the field
\ $\phi_{\epsilon}(\beta)$\
will be used
in  SGM model also.
In  the case of  \ $SU(2)$\ TM we have
\eqn\kfjg{s=\frac{1}{2},}
and the oscillator modes \ $a_m$\ satisfy the commutation relation
\eqn\ytrf{[a_{m},a_{n}]=\frac{\sinh\frac{\pi m \epsilon }{2}
\sinh\pi m \epsilon}{m}
\exp\frac{\pi |m|\epsilon }{2}
\delta_{m+n,0}.}
The operator\ \teresa\ commutes as\ \rimaqas\
and the function\ $S_{\epsilon}$\  obeys the necessary requirements.
It can be represented in  the form :
\eqn\nvb{S_{\epsilon}(\beta)=\exp(- i\epsilon s\beta)
\frac{g_{\epsilon}(-\beta)}{g_{\epsilon}(\beta)}.}
The function\ $g_{\epsilon}(\beta)$\ defines the commutator
of
the positive and negative frequency
parts \ $\phi_{\epsilon}^{\pm}(\beta)$\ of
the field \ $\phi_{\epsilon}(\beta)$\
\eqn\fhggg{\eqalign{&
[\phi_{\epsilon}^{+}(\beta_1),\phi_{\epsilon}^{-}(\beta_2)]=
-\ln g_{\epsilon}(\beta_2-\beta_1),\cr
&
g_{\epsilon}(\beta)=\exp[-\sum_{m=1}^{\infty}
\frac{\exp\frac{\pi m \epsilon }{2}}
{2 m\cosh\frac{\pi m \epsilon  }{2}}
\exp(-i m  \epsilon
\beta)].}}
The sum\ \fhggg\     converges only for \ $\Im m\beta<0$\ .
It can be   analytically continued  to the whole complex
plane by the   following:
\eqn\fdgsa{
g_{\epsilon}(\beta)=[1-\exp(-2\pi \epsilon)]^{\frac{1}{2}}
\frac{\Gamma_{\epsilon}(\frac{1}{2}+\frac{i\beta}{2\pi})}
{\Gamma_{\epsilon}(\frac{i\beta}{2\pi})},}
here
\eqn\bcfva{\Gamma_{\epsilon}(x)=[1-\exp(-2 \pi \epsilon)]^{1-x}
\prod_{k=1}^{+\infty}\frac{1-\exp(-2 \pi \epsilon k)}
{1-\exp(-2 \pi \epsilon(x+k-1))}.}
In the limit\ $\epsilon\rightarrow 0\ $\
the "quantum" \ $\Gamma  $ \ - function\ \bcfva\   becomes the usual one
and the functions\ $S_{\epsilon}(\beta)\ ,g_{\epsilon}(\beta)$\    tend
to \ $S(\beta),\ g(\beta)$\ .
Note that the constant \ $k$\ in the formula\ \yt\ is
connected with the ultraviolet regularization of the theory.
So all the  final formulas should not depend on it.

Let us consider  regularized versions of the operators
introduced  in the last section.
\eqn\kdfjf{\eqalign{&\bar\phi_{\epsilon}(\beta)=
\phi_{\epsilon}(\beta+i\frac{\pi}{2})+
\phi_{\epsilon}(\beta-i\frac{\pi}{2}),\cr
&V_{\epsilon}(\beta)=\exp(i\phi_{\epsilon} (\beta)),\cr
&\bar V_{\epsilon}(\beta)=\exp (-i\bar \phi_{\epsilon} (\beta)),\cr
&{\cal X}_{\epsilon}=\eta^{-1}_{\epsilon}
\int_{-\frac{\pi}{\epsilon}}^{\frac{\pi}{\epsilon}} \frac{d\gamma}{2\pi}
\bar V_{\epsilon}(\gamma) .}}

The following Proposition explains  our choice of the  regularization.

{\bf Proposition 3.}

The operators
\eqn\terw{\eqalign{&Z_{\epsilon+}(\beta)=\exp(\frac{i\epsilon\beta}{4})
V_{\epsilon}(\beta)
,\cr
&Z_{\epsilon-}(\beta)=
i\ \exp(-\frac{i\epsilon\beta}{4})
\big[\exp\frac{\pi
\epsilon}{4}\ {\cal X}_{\epsilon}V_{\epsilon}(\beta)+
\exp(-\frac{\pi\epsilon}{4})\ V_{\epsilon}(\beta)
{\cal X}_{\epsilon}\big],}}
obey the Zamolodchikov-Faddeev commutation relations\ \uytras\  with the
\ $S$-matrix:
\eqn\jfhg{\eqalign{&S_{++}^{++}(\beta)=S_{--}^{--}(\beta)=
S_{\epsilon}(\beta),\cr
&S_{+-}^{+-}(\beta)=S^{-+}_{-+} (\beta)=S_{\epsilon}(\beta)\frac{\sinh\frac
{i\epsilon\beta}{2}}{\sinh\frac{i\epsilon(i \pi-\beta)}{2}},\cr
&S_{+-}^{-+}(\beta)=S^{+-}_{-+}(\beta)=-S_{\epsilon}(\beta)
\frac{\sinh\frac{\pi \epsilon}{2}}{\sinh\frac{i\epsilon(i \pi-\beta)}{2}}.}}
The operator product
$$Z_{\epsilon a}(\beta_2)Z_{\epsilon b}(\beta_1) $$
has a simple pole at the point\ $\beta_2=\beta_1+i\pi$\ .
If the constant \ $\eta_{\epsilon}$\ reads
\eqn\kfjg{\eta_{\epsilon}=\big[\frac{2}{i\epsilon}\sinh\frac{\pi\epsilon}{2}
\big]^{\frac{1}{2}},}
then the residue is given by
\eqn\lgkjh{i\ Z_{\epsilon a}(\beta_2)Z_{\epsilon b}(\beta_1)=
\frac{\delta_{a+b,0}}{\beta_2-\beta_1-i\pi}+...\ \ \ .\ \spadesuit}

\subsec{Second "screening charge"}

As discussed already, we regard
the operator \ $ {\cal X}$\  as the
generator \ $\pi_Z[X^-]$\  of the quantum algebra
\ $U_{-1}(sl(2))$\ .  Let us  now define the  actions
of other generators.
In order  to do this, we have to introduce a set
of new fields and vertex operators.
First of all, consider the fields
\eqn\hfgdl{\eqalign{&\phi'_{\epsilon}(\alpha)=
-\sqrt{s'}(Q-\epsilon\alpha P)-
\sum_{m\ne 0}\frac{a'_m}{i\sinh( \pi m\epsilon)}\exp(i m \epsilon \alpha),
\cr
&\bar\phi'_{\epsilon}(\alpha)=
\phi'_{\epsilon}(\alpha+i\frac{\pi}{2})+
\phi'_{\epsilon}(\alpha-i\frac{\pi}{2}).}}
In the case of  \ $SU(2)$\ TM  the parameter\ $s'$\  is given by
\eqn\hfgg{s'=s=\frac{1}{2}}
and the normal modes \ $a'_m$\ satisfy the commutation relation:
\eqn\ytruf{[a'_{m},a'_{n}]=\frac{\sinh\frac{\pi m \epsilon }{2}
\sinh\pi m \epsilon}{m}
\exp(-\frac{\pi |m|\epsilon }{2})
\delta_{m+n,0}.}
They are simply connected with the oscillators\ $a_m$\ \ytrf

\eqn\gdfs{a'_{m}\exp\frac{\pi |m|\epsilon }{4}
=a_{m}\exp(-\frac{\pi |m| \epsilon }{4}).}

Define the   operators\ $V'_{\epsilon}(\alpha),\bar V'_{\epsilon}(\alpha)
$\ and\ ${\cal X}'_{\epsilon}$\ by analogy with\ \kdfjf
\eqn\mnb{\eqalign{&V'_{\epsilon}(\alpha)=\exp(i\phi'_{\epsilon} (\alpha)),\cr
&\bar V'_{\epsilon}(\alpha)=\exp (-i\bar \phi'_{\epsilon} (\alpha)),\cr
&{\cal X}_{\epsilon}=\eta^{'-1}_{\epsilon}
\int_{-\frac{\pi}{\epsilon}}^{\frac{\pi}{\epsilon}} \frac{d\delta}{2\pi}
\bar V'_{\epsilon}(\delta) .}}
By working out the operator product expansion, one may check that
\ ${\cal X}_{\epsilon},{\cal X}'_{\epsilon}$\
and\ $P$\ satisfy the commutation
relations:
\eqn\hdgs{\eqalign{&[P,{\cal X}_{\epsilon}]=-\sqrt2 {\cal X}_{\epsilon},\cr
&[P,{\cal X}'_{\epsilon}]=\sqrt2 {\cal X}'_{\epsilon}, \cr
&[{\cal X}'_{\epsilon},{\cal X}_{\epsilon}]=\frac{\sinh\frac{\sqrt2}{2}
\pi\epsilon P}{\sinh\frac{\pi\epsilon}{2}}.}}
As  follows from the  explicit  form
of  \ $Z_{\epsilon a}(\beta)$\
\terw\ ,  the coproduct for the  quantum algebra \ \hdgs\ reads
\eqn\ousd{\eqalign{&\Delta(P)=1\otimes P +P \otimes 1,\cr
&\Delta({\cal X}_{\epsilon})={\cal X}_{\epsilon}\otimes 1-
\exp(-\frac{\sqrt2}{2} P\pi\epsilon)\otimes {\cal X}_{\epsilon} ,\cr
&\Delta({\cal X}'_{\epsilon})={\cal X}'_{\epsilon}\otimes
\exp(\frac{\sqrt2}{2} P\pi\epsilon)-
1\otimes {\cal X}'_{\epsilon} .}}
If we conjecture the  identification
\eqn\fdret{{\cal X}'=\pi_Z[X^+],\ \ { \cal X}=\pi_Z[X^-],\ \ P=\pi_Z[H],}
then the commutation relations\ \hdgs\
and the coproduct\ \ousd\ become
\ \tte,\ \ou,\ \khjfg\ in the limit\ $\epsilon\to 0$\ .

\subsec{ Fock decomposition for  \ $SU(2)$\ TM }

At the conclusion of this section let us discuss
a structure of the regularized space\ $\pi_Z^{\epsilon}$\
for  \ $SU(2)$\ TM.
Define the highest vector\ $|p>$\ by the
system of   equations
\eqn\jghfd{\eqalign{a_m| p>=0,\ m>0,\cr
P| p>= p| p>,}}
where \ $ p$\ is some number.
Fock space \ $F_p$\ is generated by  all possible
vectors:
\eqn\hfgdp{a_{-m_{1}}...a_{-m_{n}}| p>,\  m_1,...m_n>0,}
and it is an irreducible representation
of the algebra\ \rimaq\ .
Any representation\ $\pi$\ of the universal enveloping algebra
generated by operators\ $a_m,\ P,\ Q$\ can be decomposed
into   a  direct sum of  Fock spaces
\eqn\hfgdsx{\pi=\oplus F_p.}
Here the parameter \ $p$\ specifies  components in the decomposition.

To describe the decomposition of the
\ $\pi_Z^{\epsilon}$\ , we have to
find admissible values of the  parameter\ $p$\ in the direct sum \ \hfgdsx.
First of all, let us determine the value \ $p_0$\ for the
vacuum state\ $|0>$\ . From the definition of\ $\pi_Z$\ ,
its vacuum vector must be
\ $U_{-1}(sl(2))$-invariant.
It is natural to assume that
the vacuum state
for  finite \ $\epsilon$\ satisfies  the
system of equations:
\eqn\rew{{\cal X}_{\epsilon}'|0>={\cal X}_{\epsilon}|0>=P|0>=0.}
The highest vector\ $|p_0>$\  with the
eigenvalue \ $p_0=0$\  is the unique solution of
\ \rew .
The operators
\ $ Z_{\epsilon a}(\beta)$\ \terw\  change an eigenvalue \ $p$\
to \ $p\pm\frac{1}{\sqrt2}$\ . Hence the space \ $\pi_Z^{\epsilon}$\
is represented by the  direct sum:
\eqn\mnkh{\pi_Z^{\epsilon}=\oplus_{l\in{\rm Z}}\  F_{\frac{l}{\sqrt2}}.}

In the space\  \mnkh\ the  action of the following  operator is well
defined:
\eqn\bcvtr{K_{\epsilon}=i\epsilon H_C-i\frac{\sqrt2}
{4}\epsilon P,}
Here
$$H_C= \frac {P^2}{2}+\sum_{m=1}^{+\infty}
\frac{m^2}{\sinh\frac{\pi m \epsilon} {2}\ \sinh\pi\epsilon m }
a'_{-m}a_m .$$
It is easy to check that the operator \ $K_{\epsilon}$\
generates an infinitesimal shift of the variable \ $\beta$\  for
\ $Z_{\epsilon a}(\beta)$\ . Hence we can consider the operator
\ $K$\ \bcvcl\  acting in the space\ $\pi_Z$\ as the limit
of \ $K_{\epsilon}$\ when\ $\epsilon\to 0$.

\newsec{Additional structures in the space\ $\pi_Z$\ }
In this section I shall introduce
all operators  which are necessary for
reconstruction of form-factors
in  \ $SU(2)$\ TM . There are analogical
structures in  SGM. So,
statements will be formulated in  universal forms.

\subsec{Principle  Theorem}

Let us define the following  operators in the space\ $\pi_Z$\ .
\eqn\tersad{\eqalign{&Z'_+(\alpha)=V'(\alpha)\ ,\cr
&Z'_-(\alpha)=i({\cal X}'V'(\alpha)+V'(\alpha)
{\cal X}')\ .}}
Principal Theorem summarizes  properties of the algebra
generated by  the operators\ $Z_a(\beta),\ Z'_a(\alpha)$  .

{\bf Theorem }

1. The operators \ $Z_a(\beta),\ Z'_a(\alpha)$\  satisfy the
commutation relations:
\eqn\vbcj{\eqalign{&Z_a(\beta_1)Z_b(\beta_2)=S_{ab}^{cd}(\beta_1-\beta_2)
Z_d(\beta_2) Z_c(\beta_1),\cr
&Z'_a(\alpha_1)Z'_b(\alpha_2)=R_{ab}^{cd}(\alpha_1-\alpha_2)
Z'_d(\alpha_2) Z'_c(\alpha_1),\cr
&Z_a(\beta)Z'_b(\alpha)=a\ b\ \tan(\frac{\pi}{4}+i\frac{\beta-\alpha}{2})
Z'_b(\alpha) Z_a(\beta).}}
In the case of  \ $SU(2)$\ TM the matrix \ $S_{ab}^{cd}(\beta)$\
is given by\ \jvhbf\  and
\eqn\hfgd{R_{ab}^{cd}(\alpha)=-S_{ab}^{cd}(-\alpha).}

2. The singular part of the operator product
$$Z_a(\beta_2)Z_b(\beta_1)$$
considered as a function of the complex  variable
\ $\beta_2$\ for real \ $\beta_1$\
in the upper half plane $\Im m \beta_2\geq 0$\ ,
contains  only one simple pole with
the residue

\eqn\alsa{  i\ Z_a(\beta_2)Z_b(\beta_1)=
\frac{\C_{ab}}{\beta_2-\beta_1-\pi i}+...\ \ \  .}

3. The operator product
$$Z'_a(\alpha_2)Z'_b(\alpha_1)$$
considered as a function of the complex variable\ $\alpha_2$\ for
real\ $\alpha_1$\  is  regular  for \ $\Im m\alpha_2\geq-\pi $\
and
\eqn\hfgda{\eqalign{&\C_{ab}Z'_a(\alpha+i\pi) Z'_b(\alpha)=i,\cr
&Z'_a(\alpha-i\pi) Z'_b(\alpha)=i\ \C_{ab}\ .}}

4. The following combination
$$\frac{\Gamma(\frac{3}{4}+i\frac{\beta-\alpha}{2 \pi})}
{\Gamma(\frac{1}{4}+i\frac{\beta-\alpha}{2 \pi})}Z'_a(\alpha)Z_b(\beta)$$
considered as a function of the variable\ $\beta$\ is  regular
in the whole complex plane.\ $ \spadesuit$

The proof of the theorem is close to the proofs of Propositions 1,2
from Sec. 5.4 .  It  is based on
explicit expressions for operator products of the
vertices
$$V(\beta),\bar V(\gamma),V'(\alpha),\bar V'(\delta).$$
To get them, it is necessary to apply  the oscillator  representation
from  Sec. 6 and then consider the limit\ $\epsilon\to 0$\ .
The result of the calculations is presented in the Appendix 2.

\subsec{Symmetry algebra of the space of local operators}

As it follows from discussion in  Sec.3, the problem of a  description
of the  space of local operators is  reduced to finding
operators\ $\Lambda(O)\in {\rm End}[\pi_Z]$\
satisfying\ \lor\ . The operators \ $Z'_a(\alpha)$\
allow us to solve this problem as follows.

In the case of  \ $SU(2)$\ TM let us introduce the set of
operators:
\eqn\jhfx{T(\alpha)=\frac{1}{i} \C_{ab}Z'_a(\alpha+i\frac{\pi}{2})
\partial_{\alpha}Z'_b(\alpha-i\frac{\pi}{2}),}
\eqn\hfgdt{\eqalign{\Lambda_m(\alpha)=
\frac{i}{\eta'}\left[\matrix{\frac{1}{2}&\frac{1}{2}&1\cr
\frac{a}{2}&\frac{b}{2}&m}\right]_{-1} Z'_a(\alpha+i\frac{\pi}{2})
Z'_b(\alpha-i\frac{\pi}{2}).}}
Note that due to the  formula \  \hfgda\ the  operators
\ \jhfx ,\ \hfgdt\
can be represented in the
equivalent forms:
\eqn\nbvsx{T(\alpha)=\frac{i}{2}\C_{ab}Z'_a(\alpha-i\frac{\pi}{2})
\partial_{\alpha}Z'_b(\alpha+i\frac{\pi}{2})+const,}
\eqn\pogi{\eqalign{\Lambda(\alpha)=\frac{\eta'}{i}
\left[\matrix{\frac{1}{2}&\frac{1}{2}&1\cr
\frac{a}{2}&\frac{b}{2}&m}\right]_{-1}Z'_a(\alpha-i\frac{\pi}{2})
\partial_{\alpha}Z'_b(\alpha+i\frac{\pi}{2}),}}
where the irrelevant constant reads
$$const=\frac{i}{2\pi}(1-4 \ln 2).$$
We shall need also the bosonic forms of
$\ \Lambda_m(\alpha)$\
\eqn\ghfgd{\eqalign{
&\Lambda_{1}(\alpha)=\tilde V'(\alpha)\ ,\cr
&\Lambda_0(\alpha)=\frac{i}{\sqrt2}\left[{\cal X}'\tilde V'(\alpha)
-\tilde V'(\alpha){\cal X}'\right]\ ,\cr
&\Lambda_{-1}(\alpha)=\frac{1}{2}\left[{\cal X}^{'2}\tilde V'(\alpha)
-2{\cal X}'\tilde
V'(\alpha){\cal X}'+
\tilde V'(\alpha){\cal X}^{'2}\right].}}
Here the new vertex operator
\eqn\mnn{\tilde V'(\alpha)=\exp(i \bar\phi'(\alpha)).}
is introduced.

The  essential properties  of the operators\ $T(\alpha),\ \Lambda_m(\alpha)
$\ are  direct consequences of  Principle Theorem;

{\bf Proposition 4}

1. The operators \ $T(\alpha),\ \Lambda_m(\alpha)$\ generate
the quadratic algebra:
\eqn\gdfs{\eqalign{\Lambda_{m_1}(\alpha_1)\Lambda_{m_2}(\alpha_2)=&
{\cal R}_{m_1m_2}^{m_3m_4}(\alpha_1-\alpha_2) \Lambda_{m_4}(\alpha_2)
\Lambda_{m_3}(\alpha_1)+\cr
&+{\cal A}_{m m_1m_2}(\alpha_1-\alpha_2)(\Lambda_{m}(\alpha_1)+
\Lambda_m(\alpha_2) )+{\cal A}_{m_1m_2}(\alpha_1-\alpha_2),}}
\eqn\nvbb{\eqalign{\big[\Lambda_m(\alpha_1),T(\alpha_2)\big]=
-{\cal A}_{m m_2m_1}&(\alpha_1-\alpha_2)
\Lambda_{m_2}(\alpha_2)\Lambda_{m_1}(\alpha_1)+\cr
&+{\cal A}(\alpha_1-\alpha_2)\Lambda_{m}(\alpha_1)+
{\cal B}(\alpha_1-\alpha_2)\Lambda_{m}(\alpha_2),}}
\eqn\bcvd{-\eta^{'2}\big[T(\alpha_1),T(\alpha_2)\big]=
{\cal A}_{m_2m_1}(\alpha_1-\alpha_2)
\Lambda_{m_2}(\alpha_2)\Lambda_{m_1}(\alpha_1)+{\cal C}(\alpha_1-\alpha_2).}
One can find explicit expressions for the  structure functions
\ ${\cal R},{\cal A},{\cal B},{\cal C}$\
in the   Appendix 3.

2. The operators \ $T(\alpha),\ \Lambda_m(\alpha)$\  obey the
following commutation relations with \ $Z_a(\beta):$
\eqn\kiu{
\Lambda_m(\alpha)Z_a(\beta)=(-1)^m Z_a(\beta)\Lambda_m(\alpha),}
\eqn\yfdrs{\big[T(\alpha),Z_a(\beta)\big]=\partial_{\alpha}
\ln s(\alpha-\beta)
Z_a(\beta),}
here
$$s(\alpha)=\coth\frac{\alpha}{2}\ .$$

3. The combination
\eqn\fds{(\alpha-\beta)\Lambda_m(\alpha)Z_a(\beta)}
considered as a function of the variable\ $\beta$\ is a regular one
in the hole complex plane.\ $\spadesuit$

It is useful to compare \ \kiu\  with the formula
\ \lor .
The operators\ $\Lambda_m(\alpha)$\ satisfy the
proper  commutation relations with \ $Z_a(\beta)$\ .
In the next section it will be explained that they
define generating
functions for form-factors.
Here I wish to
note only that the sign factor in the formula\ \kiu\
is connected with the mutual locality indices\ \hdgfs.

To clarify the meaning of the operator \ $T(\alpha)$\ let
us return to the general  consideration from  Sec. 3.
Note that if  \ $O(x)$\ is  a local  operator
then its commutators with IM\ \bcv
\eqn\hsghfd{O(x,s)=\big[O(x),I_s\big]}
are also  local operators. The generating function for
the form-factors of \ \hsghfd\
reads
\eqn\jdhgfg{F_{a_1...a_n}(\alpha|\beta_1,...\beta_n)=
\sum_{k=1}^n\p_{\alpha}\ln s(\alpha-\beta_k)
F_{a_1...a_n}(\beta_1,...\beta_n),}
where the function \ $s(\alpha)$\  is defined by the equation\ \gdfsd\
and \ $F_{a_1...a_n}(\beta_1,...\beta_n)$\ is the
form-factor of the operator\ $O(x)$.
Suppose that the function \ $F_{a_1...a_n}(\beta_1,...\beta_n)$\ can
be represented by the formula\ \tr\  with a
some  \ $\Lambda(O)\in {\rm End }[\pi_Z]$\ . Then
the generating function\ \jdhgfg\
is also given  by the trace
\eqn\gdfasd{F_{a_1...a_n}(\alpha|\beta_1,...\beta_n)
=\Tr_{\pi_Z}\big[\exp(2\pi i K)
\Lambda(O,\alpha)Z_{a_n}(\beta_n)...Z_{a_1}(\beta_1)\big]\ ,}
where
\eqn\hdgdfas{\Lambda(O,\alpha)=\Lambda(O)T(\alpha)-T(\alpha+2 i\pi)
\Lambda(O),}
and the operator \ $T(\alpha)$\ satisfies the equation\ \yfdrs.

Thus the algebra\ \gdfs -\bcvd\ determines a  structure of the space
of local operators in the theory.

\newsec {Trace calculations}
In this section we will get the integral representation
for form-factors in  \ $SU(2)$\ TM.

Consider the following functions:
\eqn\hgdds{\eqalign{{\cal F}_{a_1...a_n}^{m_1...m_k}(\alpha_1,...\alpha_k|&
\beta_1,...\beta_n)=\cr
&=Tr_{\pi_Z}\big[\exp(2\pi i K) \Lambda_{m_k}(\alpha_k)
...\Lambda_{m_1}(\alpha_1)Z_{a_n}(\beta_n)...Z_{a_1}(\beta_1)\big].}}
Due to the boson representation for the  operators \ $Z_a(\beta)$\
and \ $\Lambda_m(\alpha)$\ ,
they  are given by combinations
of multiple contour integrals. The
integrands are functions like
\eqn\tdrsas{\eqalign{R(\alpha_1&,...\alpha_k|\delta_1,...\delta_p|
\beta_1,...\beta_n|\gamma_1,...\gamma_r)=\cr
&=Tr_{\pi_Z}\big[\exp(2\pi i K)\tilde V(\alpha_k)...\tilde V(\alpha_1)
\bar V(\delta_p)...\bar V(\delta_1)V(\beta_n)...V(\beta_1)\bar V(\gamma_r)...
\bar V(\gamma_1)\big]}}
Here
 \ $\ \{\delta_i\},\ \{\gamma_j\}$\  are  variables of integrations.
To treat the traces\ \tdrsas\ it is useful to consider
the space\ $\pi_Z$\ as the limit\ \lgjkh
\eqn\tdrsasqs{R(\alpha_1,...)=\lim_{\epsilon\to 0}Tr_{\pi^{\epsilon}_Z}
\big[\exp(2\pi i K_{\epsilon})\tilde V_{\epsilon}(\alpha_k),...\big].}
According to the formula\ \mnkh\
the trace over \ $\pi^{\epsilon}_Z$\
is a product  of the  traces  over Fock module\ $F$(\ $Tr_F$\ )
and  the space of zero modes (\ $Tr_0$\ ).
First, let us  consider the second one.
\eqn\kfhgh{\eqalign{Tr_0\big[\exp(2\pi i K_{\epsilon})&\tilde V_{\epsilon}
(\alpha_k)...\big]=\cr
&=\delta_{n-2 r+2 p-2k,0}
f\big(\sum_{j=1}^n\beta_j
-2\sum_{j=1}^r\gamma_j+2\sum_{j=1}^p\delta_j
-2\sum_{j=1}^k\alpha_j\big) .}}
Here
$$f(\beta)=\sum_{l=-\infty}^{+\infty}\exp\big(-\frac{\pi\epsilon l^2}{2}-
\frac{i \epsilon l}{2}(\beta+\pi i)\big).$$

The limit\ $\epsilon\to 0$\
can be calculated directly if we  apply  the Poisson formula:
\eqn\hdgf{\sum_{n=-\infty}^{+\infty}u(n\delta)=
\delta^{-1}\sum_{n=-\infty}^{+\infty} v(2\pi n/\delta),}
where
$$v(k)=\int_{-\infty}^{+\infty}d x\ \exp(i k x) u(x).$$
After a little algebra one can  find that the function \ $f(\beta)$\ \kfhgh\
equals to
(infinite) constant. Hence we can set up
\eqn\kfjghop{Tr_0\big[\exp(2\pi i K)\tilde V(\alpha_k)...\big]=
\delta_{n-2 r+2 p-2k,0}\ .}

The calculation  of  the trace over Fock module is
simplified by the technique
of Clavelli and Shapiro\ \cls ,
\japthree . Their prescription is as follows:
Introduce a copy of bosons\ ${b_n}$\  satisfying \ $[a_m,b_n]=0$\ and
the same commutation relations as the\ $ a_n$\ . Let
\eqn\jfhg{\tilde a_m=\frac{a_m}{1-\exp(-2\pi m\epsilon)}+b_{-m}\ \ (m>0)\ ,\ \
\tilde a_m=a_m+\frac{b_{-m}}{\exp(-2\pi m\epsilon)-1}\ \ (m<0)\ .}
For a linear operator \ ${\cal O}(\{a_n\})$\ on the Fock space \ $F[a]$\ ,
let\ $\tilde{\cal O}={\cal O}(\{\tilde a_n\})$\  be the operator on
\ $ F[a]\otimes F[b] $\ obtained by substituting\  $\tilde a_{m}$\ for
\  $a_{m}$\ . We have then
\eqn\fsda{Tr_{F}\big[\exp(2\pi i K_{\epsilon}) {\cal O}\big]
=\frac{<0|\tilde
{\cal O}|0>}{\prod_{m=1}^{\infty}(1-\exp(-2\pi m\epsilon))},}
where \ $<0|\tilde
{\cal O}|0>$\  denotes the usual expectation value with respect to the
Fock vacuum \ $|0>\in F[a]\otimes F[b]\ ,<0|0>=1 $\ .

Here is the final result of calculations of the functions\ \tdrsas:
\eqn\nvbc{\eqalign{&R(\alpha_1,...\alpha_k|\delta_1,...\delta_p|
\beta_1,...\beta_n|\gamma_1,...\gamma_r)=\cr
& = {\cal C}_1^{-\frac{n}{2}} {\cal C}_2^{\frac{n+2r}{4}} {\cal C}_2^
{'\frac{p+k}{2}}
 \ 2^{-r-\frac{n}{2}}i^{\frac{n}{2}+ p+ k}\eta^r\eta^{'-p-k}
T_0(\{\alpha_i\},\{\delta_i\},\{\beta_i\},\{\gamma_i\})
\times\cr
&\prod_{ 1\le i<j\le n} G(\beta_i-\beta_j) \prod_{ 1\le i<j\le r} \bar
G(\gamma_i-\gamma_j) \prod_{ \scriptstyle 1\le i\le r\atop
\scriptstyle 1\le j\le n} W(\gamma_i-\beta_j)\times\cr
&\prod_{ 1\le i<j\le k}\bar G'(\alpha_i-\alpha_j) \prod_{ 1\le i<j\le p}
\bar G'(\delta_i-\delta_j)\prod_{ \scriptstyle 1\le i\le p\atop
\scriptstyle 1\le j\le k}\bar G^{'-1}(\delta_i-\alpha_j)\times\cr
&\prod_{ \scriptstyle 1\le i\le n\atop
\scriptstyle 1\le j\le k} U^{-1}(\beta_i-\alpha_j)
\prod_{ \scriptstyle 1\le i\le n\atop
\scriptstyle 1\le j\le p} U(\beta_i-\delta_j)
\prod_{ \scriptstyle 1\le i\le r\atop
\scriptstyle 1\le j\le k} \bar H^{-1}(\gamma_i-
\alpha_j)
\prod_{ \scriptstyle 1\le i\le r\atop
\scriptstyle 1\le j\le p} \bar H(\gamma_i-
\delta_j)\ .}}
For the case of \ $SU(2)$\ TM the functions and constants in the
formula\ \nvbc\  have the following explicit expressions:
\eqn\jdhfplkk{\eqalign{&
T_0(\{\alpha_i\},\{\delta_i\},\{\beta_i\},\{\gamma_i\})=
\delta_{n-2 r+2 p-2k,0}
\ ,\cr
&G(\beta)= i\  {\cal C}_1 \sinh \frac{\beta}{2}
\exp\big[\int_{0}^{+\infty}\frac{d t} {t} \frac {\sinh^2t (1-i \frac{\beta}
{\pi}) \exp(- t)}{\sinh 2 t \cosh t}\big]\ ,\cr
&W(\beta)= (\pi ^3{\cal C}_2)^{-\frac{1}{2}} \
\Gamma(-\frac{1}{4}+i\frac{\beta}{2\pi})
\Gamma(\frac{3}{4}-i\frac{\beta}{2\pi})\ ,\cr
&\bar G(\beta)=-\frac{{\cal C}_2}{4}\  (\beta+i\pi) \sinh\beta\ , \cr
&\bar G'(\alpha)=-{\cal C}'_2\ \frac{\sinh\alpha}{\alpha+i\pi}\ , \cr
&U(\alpha)=i\ \sinh\frac{\alpha}{2}\ ,\cr
&\bar H(\alpha)=-\frac{2}{\cosh\alpha}\ , \cr
&{\cal C}_1=
\exp\big[-\int_{0}^{+\infty}\frac{d t} {t} \frac {\sinh^2\frac{t }
{2} \exp(- t)}{\sinh 2 t \cosh t}\big]\ , \cr
&{\cal C}_2={\cal C}_2^{'-1}=\frac{\Gamma^4(\frac{1}{4})}{4 \pi^3}\  .}}
Now, to complete the construction  of integral representations
for  the functions\ \hgdds \ , we have to describe the
contours of integrations.
The rules are:
If all\ $\{\beta\}_i,\{\alpha\}_i$\ are real   then  contours of
integrations over the variables\ $\{\delta\}_j,\{\gamma\}_j$\  lie
in the strip $-i\pi<\{\delta\}_j,\{\gamma\}_j<i\pi$
exactly in the   same way  as in the integral representation for the
vacuum averages
$$<0|\Lambda_{m_k}(\alpha_k)
...\Lambda_{m_1}(\alpha_1)Z_{a_n}(\beta_n)...Z_{a_1}(\beta_1)|0>.$$
Let me recall that  contours  for vacuum averages
are taken according to  the definition  of  the action of the
operators\ ${\cal X},{\cal X}'$\ (see the Sec. 5.3).

Now it is useful to consider  examples of
\ \hgdds \ .
I  calculated explicitly the  simplest ones.
\eqn\kdjhfh{{\cal F}_{ab}(\beta_1,\beta_2)=0,\ \ \ \ \
{\cal F}^{m}(\alpha)=0,}
\eqn\mdhf{\eqalign{
{\cal F}^{m}_{ab}(\alpha|\beta_1,\beta_2)=&\frac{\eta'}{2\pi i}
\left[\matrix{\frac{1}{2}&\frac{1}{2}&1\cr
\frac{b}{2}&\frac{a}{2}&m}\right]_{-1}
{\cal C}_1^{-1}
\frac{G(\beta_1-\beta_2)}
{\sinh\frac{\beta_1-\alpha}{2} \sinh\frac{\beta_2-\alpha}{2}}.}}
Let us analyze these formulas.
One should expect that the functions  \ $ {\cal F}_{ab}(\beta_1,\beta_2)$\
are  two particle form-factors of  the
unit operator. Hence it must be zero for
real\ $\beta_1-\beta_2$\ .  Then,
we can see that  \ ${\cal F}_{ab}^{m}(\alpha|\beta_1,\beta_2)$\ ,
considered
as functions of\ $\exp(\alpha)$\
admit   decompositions
into the series in  the neighborhood of the points \ $\exp(\alpha)=0$\ and
\ $ \exp(\alpha)=\infty$\  :
\eqn\kfkj{\eqalign{&{\cal F}_{ab}^{m}(\alpha|\beta_1,\beta_2)=
\sum_{s=1}^{+\infty}
\exp(s\alpha) F_{ab}^{m}(s|\beta_1,\beta_2)\ , \cr
&{\cal F}_{ab}^{m}(\alpha|\beta_1,\beta_2)=\sum_{s=-1}^{-\infty}
\exp(s\alpha)F_{ab}^{m}(s|\beta_1,\beta_2).}}
The coefficients  \ $F_{ab}^{m}(s|\beta_1,\beta_2)$\
are  form-factors of operators which are \ $SU(2)$\  vectors and
have  Lorentz spins \ $s$\ . Moreover,
\ $F_{ab}^{\pm}(s|\beta_1,\beta_2)$\ and
 \ $F_{ab}^{0}(s|\beta_1,\beta_2)$\
correspond to operators
which are respectively semilocal and local
with respect to the "elementary" field of \ $SU(2)$\ TM. This follows from
the commutation relations\ \kiu\ . So, we  conjecture that
the functions \  $F_{ab}^{m}(s|\beta_1,\beta_2)\ ,(s=\pm1,\ m=0,\pm) $\
are the form-factors of the currents\
$ -\frac{2}{ \eta'} [s\ J^{m}_{t}(x)+
J^{m}_{x}(x)]$\ \kar ,\ \ki .
Note that  the constant of  normalization is fixed by the equations\ \mnj\ .

{}From this simple example we have learned
that \ \ ${\cal F}_{ab}^{m}(\alpha|\beta_1,\beta_2)$\
are the generating
functions for form-factors of local operators in the theory.
As a matter of fact
it is the general property of the
functions\  \hgdds \ .
Indeed,
using
Proposition 4 from Sec. 7.2,
we can show  that  any function\ \hgdds \
has the form:
\eqn\hgs{\eqalign{{\cal F}_{a_1...a_n}^{m_1...m_k}&(\alpha_1,...\alpha_k|
\beta_1,...\beta_n)=\cr
&=A^{m_1...m_k}(\alpha_1,...\alpha_k)B_{a_1...a_n}(\beta_1,...\beta_n)
C(\alpha_1,...\alpha_k|
\beta_1,...\beta_n)\ ,}}
where
$$C(\alpha_1,...,\alpha_j+2\pi i,...\alpha_k|\beta_1,...\beta_n)
=C(\alpha_1,...,\alpha_j,...\alpha_k|\beta_1,...\beta_n)\ ,\ j=1,...k  $$
Scalar periodic functions \
$C(\alpha_1,...\alpha_k|\beta_1,...\beta_n)$\  admit,
in the neighborhood of the
points\ $\exp(\alpha_j)=0,\infty$,
decompositions
into the series.
Hence  the functions\ \hgs\ can be represented as follows
\eqn\hsgdj{\eqalign{{\cal F}_{a_1...a_n}^{m_1...m_k}&(\alpha_1,...\alpha_k|
\beta_1,...\beta_n)=\cr
&=\sum_{\{s_j\}}F^{'m_1...m_k}(\alpha_1,...\alpha_k|\{s_j\})
F_{a_1...a_n}(\{s_j\}|\beta_1,...\beta_n) \exp(s_1\alpha_1)...
\exp(s_k\alpha_k)  .}}
The coefficients\
$ F_{a_1...a_n}(\{s_j\}|\beta_1,...\beta_n)$\   satisfy
the axioms (1-5) from Sec. 3, so they are  form-factors
of local operators in the theory. One can expect that this huge set of
functions is a  general solution of the Riemann-Hilbert problem
(1-5) for  \  $SU(2)$\ TM.

\newsec{ Free field representation for the Sine-Gordon model}

In this section the representation\ $\pi_Z $\ for
SGM will be  investigated.
The main steps of the construction have been already  discussed
for the case of  \ $SU(2)$\ TM,
so we shall focus only on essential differences.

\subsec{Bosonization of the Zamolodchikov-Faddeev algebra}

Let us begin  with a definition of the
space\ $\pi^{\epsilon}_Z$\ .
Consider the set of oscillators\ $a_m$\ , which satisfy the commutation
relation:
\eqn\ytrfsd{[a_{m},a_{n}]=\sinh\frac{\pi m \epsilon }
{2}\sinh\pi m \epsilon \frac{\sinh\frac{\pi m \epsilon }{2}(\xi+1)}
{m \sinh\frac{\pi m\epsilon  }{2}\xi}\delta_{n+m,0}.}
It is also convenient  to introduce  the set of "dual" oscillators
\ $a'_m$ \ connected with \ $a_m$\ as follows:
\eqn\kjbv{a'_m \sinh\frac{\pi m\epsilon } {2}(\xi+1)=
a_m\sinh\frac{\pi m \epsilon }{2}\xi.}
They obey the commutation relations:
\eqn\jfgv{[a'_{m},a'_{n}]=\sinh\frac{\pi m \epsilon }
{2}\sinh\pi m \epsilon \frac{\sinh\frac{\pi m \epsilon }{2}\xi}
{m \sinh\frac{\pi m\epsilon  }{2}(\xi+1)}\delta_{n+m,0}.}

Note the  duality between the operators
\ $a_m$\ and\ $a'_m$\ ;  \ $Z_2$\ -  transformation
\eqn\gdfsaq{\xi\rightarrow -1-\xi}
transforms the commutation relation \ \jfgv\
into \ \ytrfsd\   and vice versa.
As we shall see below, this duality has the
same nature as the "$\alpha_+\leftrightarrow\alpha_-$"one in the
minimal model of  2D CFT
\ref\bpz{A.A. Belavin, A.M. Polyakov and A.B. Zamolodchikov,
Nucl. Phys. B241 (1984) 333.},
\dot .

Using the oscillators\ $a_m,a'_m$\ , we  define the fields
\ $\phi(\beta)$\ and\ $\phi'(\alpha)$\ in the same
way as have been done for  $SU(2)$\ TM\ \teresa,\ \hfgdl\  . One should
point out only that the parameters \ $s$\ and\ $s'$\ must
be chosen as
\eqn\jfhfg{s=\frac{\xi+1}{2\xi}\ ,\ \  \  s'=\frac{\xi}{2(\xi+1)}.}
It may be useful to recall that  \ $s$\ is connected with an asymptotic
behavior of the two particle \ $S$-matrix\ \kgjh .

Introduce the operators
$$V_{\epsilon}(\beta),\bar V_{\epsilon}
(\gamma),V'_{\epsilon}(\alpha),\bar V'_{\epsilon}(\delta),{\cal X}_{\epsilon},
{\cal X}_{\epsilon}'$$
by the formulas\ \kdfjf,\ \mnb\  .
Using \ ${\cal X}_{\epsilon}$\ and \ ${\cal X}_{\epsilon}' $\ ,
the vacuum state can be specified  as follows:
\eqn\hgdfs{{\cal X}_{\epsilon}|0>={\cal X}_{\epsilon}'|0>=0.}
Their unique solution is the highest vector\  $|p_0>$\ with
\eqn\ldkfjg{p_0=\frac{\alpha_++ \alpha_-}{\sqrt2}  }
where we have conveniently used the  notations:
\eqn\djhfp{\alpha_+=-\alpha_-^{-1}=\sqrt{\frac{\xi+1}{\xi}}.}
The vertices \ $V_{\epsilon}(\beta),\bar V_{\epsilon}(\gamma),
V'_{\epsilon}(\alpha),\bar V'_{\epsilon}(\delta)$\
shift an eigenvalue\  $p$\  of the operator \ $P$\
to, respectively,
$$p+\frac{\alpha_+}{\sqrt2}\ ,\ p-\sqrt2 \alpha_+\ ,\
p+\frac{\alpha_-}{\sqrt2}\ ,\ p-\sqrt2 \alpha_-.$$
Hence  one can   expect
that  the space\ $\pi_Z^{\epsilon}$\ admits the following decomposition:
\eqn\mnkhaa{\pi_Z^{\epsilon}=\oplus_{\{l,l'\}\in{\rm Z}}\
F_{\frac{\alpha_+ l+\alpha_-l'}{\sqrt2}}.}

The operator\ $K_{\epsilon}$\  is defined by
\eqn\bcvtraksj{K_{\epsilon}=i\epsilon H_C+\frac{\alpha_++\alpha_-}
{\sqrt2} P,}
here
$$H_C= \frac {P^2-p^2_0}{2}+\sum_{m=1}^{+\infty}
\frac{m^2}{\sinh\frac{\pi m\epsilon}{ 2}\ \sinh\pi m \epsilon  }
a'_{-m}a_m .$$
It satisfies the proper
commutation relation with
 \ $V_{\epsilon}(\beta)$
\eqn\hsgdfb{\exp(-\theta K_{\epsilon})V_{\epsilon}(\beta)
\exp(\theta K_{\epsilon})=V_{\epsilon}(\beta+\theta)
\exp(-\frac{\theta}{2\xi})\ .}

The space\  $\pi_Z$\ must be considered    as the limit\ \lgjkh .
It  allows  to determine  forms of  necessary operator products.
They  are listed in the Appendix 2.

Now, we can introduce the following set of operators acting in the
space\ $\pi_Z$\ :
\eqn\terass{\eqalign{& Z_+(\beta)=\exp(-\frac{\beta}{2\xi})V(\beta),\cr
& Z_-(\beta)=\exp(\frac{\beta}{2\xi})[q^{\frac{1}{2}}{\cal X}V(\beta)-
q^{-\frac{1}{2}}V(\beta)
{\cal X}],\cr
& Z'_+(\alpha)=\exp\big(-\frac{\alpha}{2(\xi+1)}\big)V'(\alpha),\cr
& Z'_-(\alpha)=\exp\big(\frac{\alpha}{2(\xi+1)}\big)
[q^{'\frac{1}{2}}{\cal X}'V(\alpha)-q^{'-\frac{1}{2}}V'(\alpha)
{\cal X}'],}}
Here
\eqn\mvnbhf{q=\exp \ i\pi \alpha_+^2\ ,\ \ \ \
q'=\exp\ i \pi\alpha_-^2 .}

Note that
the integrals associated
with the action of the operator\ ${\cal X}$\  in \ \terass\
can be calculated explicitly at the free fermion point (\ $\xi=1$\ ).
For this case the total contour of
the integration  closes  and
the operators\ $Z_a(\beta)$\ read as:
\eqn\refd{\eqalign{& Z_+(\beta)=\exp(-\frac{\beta}{2})
\exp(i\phi(\beta)),\cr
& Z_-(\beta)=\exp\frac{\beta}{2}
\big[\exp(-i\phi(\beta+i\pi))-\exp(-i\phi(\beta-i\pi))\big].}}

Principle  Theorem from  Sec.7.1  describes the
essential properties of the operators \ \terass .
In the  case of SGM  the \ $S$-matrix defining
the commutation relations of the  two operators \ $Z_a(\beta)$\
is given by\ \jvhbfsdq\  .
At the same time the matrix \ $R_{ab}^{cd}(\alpha)$\
in the formula\ \vbcj\  has the following nontrivial elements:
\eqn\fsdq{\eqalign{&R_{++}^{++}(\alpha)=R_{--}^{--}(\alpha)=R(\alpha),\cr
&R_{+-}^{+-}(\alpha)=R^{-+}_{-+} (\alpha)=-R(\alpha)\frac{\sinh\frac{\alpha}
{\xi+1}}
{\sinh\frac{i \pi+\alpha}{\xi+1}},\cr
&R_{+-}^{-+}(\alpha)=R^{+-}_{-+}(\alpha)=R(\alpha)\frac{\sinh\frac{ i\pi}
{\xi+1}}
{\sinh\frac{i \pi+\alpha}{\xi+1}}.}}
Here the function\ $R(\alpha)$\ is represented by :
\eqn\bvvcre{R(\alpha)=\frac{\Gamma(\frac{1}{\xi+1})
\Gamma(1-\frac{i\alpha}{\pi(\xi+1)})}
{\Gamma(\frac{1}{\xi+1}-\frac{i\alpha}{\pi(\xi+1)})}
\prod_{p=1}^\infty\frac{R'_p(-\alpha)R'_p(i\pi+\alpha)}{R'_p(0)R'_p(i\pi)},}
$$R'_p(\alpha)=\frac{\Gamma(\frac{2 p}{\xi+1}+\frac{i\alpha}{\pi(\xi+1)})
\Gamma(1+\frac{2 p}{\xi+1}+\frac{i\alpha}{\pi(\xi+1)})}
{\Gamma(\frac{2 p+1}{\xi+1}+\frac{i\alpha}{\pi(\xi+1)})
\Gamma(1+\frac{2 p-1}{\xi+1}+\frac{i\alpha}{\pi(\xi+1)})}.$$

The proof of  Principle
Theorem for  SGM
is based on the ideas  given in the Appendix1. I wish to comment only
the following  commutation  relation :
\eqn\jdhfg{Z_-(\beta)Z'_-(\alpha)=\tan(\frac{\pi}{4}+
i\frac{\beta-\alpha}{2})Z'_-(\alpha)Z_-(\beta).}
To prove it we have to show that, for all vectors
\ $|u>,|v>\in \pi_Z$,
$$ |u>,|v>={\cal Z}_{a_n}...{\cal Z}_{a_1}|0>\ ,\ {\cal Z}_{a_i}=\{
Z_{a_i}(\beta_i)\  {\rm or}\ Z'_{a_i}(\alpha_i)\},$$
a  general matrix element
$$<u|\big[{\cal X}',{\cal X}\big]|v>$$
vanishes.
The  structure of the
operator products provides   exactly
this important property. Here we
have the essential difference from the
\ $SU(2)$\ TM where commutation relations
$$[{\cal X}',{\cal X}\big]=\sqrt2 P$$
holds.

\subsec{Connection with  the Feigin-Fuchs bosonization}

Now let us argue  a group theoretic meaning of the formulas\ \terass\ .
As has been mentioned in  Sec. 4 the operators
$$\hat Z_a(\beta)\equiv\exp(a\frac{\beta}{2\xi})Z_a(\beta)$$
can  be regarded  as  the basis of the fundamental representation of
\ $U_q(sl(2))$. Then the definition \ \terass\
is equivalent to  the  coproduct \ \tres\ , if the operator
\ ${\cal X}$\ is considered as
\eqn\vxcsf{{\cal X}=\pi_Z[E\ q^{-\frac{\sqrt2}{2}H}].}
The definition of the operators\ $Z_a(\alpha)$\ admits
the same interpretation. Hence  we conclude that there are
two quantum algebras\ (\ $U_q(sl(2))$\ and\ $ U_{q'}(sl(2))$\ )
in  SGM. Their quantum parameters \ $q$\ and \ $q'$\
are given by  the equations\  \mvnbhf.\
Analogous phenomena take place in 2D Conformal Field Theory
\dot ,
\ref\ffk{G. Felder, J. Froehlic and G. Keller, Comm. Math. Phys. 124
(1989) 717.}.

As a matter of fact, the algebra generated by the operators
\ $Z_a(\beta), Z'_a(\alpha)$\  is the natural generalization
of the vertex operator algebra in the minimal models. In order to
clarify this statement, let us consider
n-point vacuum functions
\eqn\tersfgf{\hat G_{a_1...a_n}(\beta_1,...\beta_n)=
<0|\hat Z_{a_n}(\beta_n)...\hat Z_{a_1}(\beta_1)|0>.}
The vacuum state\  $|0>$\ is a \ $U_q(sl(2))$-scalar,
so   matrix elements \ \tersfgf\ have the following structure:
\eqn\weds{\eqalign{&G_{a_1...a_n}(\beta_1,...\beta_n)=\cr
&\sum_{\scriptstyle\{m_k\},\{j_s\}}
\left[\matrix{\frac{1}{2}&\frac{1}{2}&j_1\cr
\frac{a_2}{2}&\frac{a_1}{2}&m_1}\right]_q\left[\matrix{\frac{1}{2}&j_1&j_2\cr
\frac{a_3}{2}&m_1&m_2}\right]_q\cdots
\left[\matrix{\frac{1}{2}&j_{n-1}&0\cr
\frac{a_n}{2}&m_{n-1}&0}\right]_q
\Im_{j_1...j_{n-1}}(\beta_1,..\beta_n).}}
Functions\ $\Im_{j_1...j_{n-1}}(\beta_1,..\beta_n)$\ are
vacuum averages of \ $U_q(sl(2))$\ scalars.
To clear up their property
it is useful to  rewrite  the variables \ $\beta_i$\ in the form
\eqn\jfhg{\beta_i=\sigma_i L}
and take a limit\ $L\rightarrow + \infty$\ .
Let us consider the commutation relation for the operators \
$\hat Z_a(\beta)$
in this limit. Using the explicit expression
for the matrix \ $\hat S(\beta)$\
\ \jvhbfsdq , \lgki  , one  can derive the formula:
\eqn\gdtrs{\hat S_{ab}^{cd}(L \sigma)\rightarrow
{\rm R}^{cd}_{ab}(q)\Theta(-\sigma)+
({\rm R}^{-1})^{dc}_{ba}(q)
\Theta(\sigma),\ L\rightarrow + \infty ,}
where
$$\Theta(\sigma)=\cases{1,&if $\sigma>0$;\cr
0,&otherwise \cr }$$
and
\ ${\rm R}^{cd}_{ab}(q)$\ are  matrix elements of the universal
R- matrix\ \drin\   in the  tensor product of   two
fundamental representations   of
\ $U_q(sl(2))$\ .
Hence the commutation relation for the operators \ $\hat Z_a(\beta)$ \ becomes
\eqn\hftsdr{\chi_{a}(\sigma_1) \chi_{b}(\sigma_2)=
[{\rm R}^{cd}_{ab}(q)\Theta(\sigma_2-\sigma_1)+
({\rm R}^{-1})^{dc}_{ba}(q)
\Theta(\sigma_1-\sigma_2)]\chi_{d}(\sigma_2)
\chi_{c}(\sigma_1).}
Here  the limit of the operator\ $\hat Z_a(L\sigma)$\ \ $(L\rightarrow
\infty)$\ is denoted as \ $\chi_a(\sigma)$.\
Now it is easy to see that   functions
$$\Im_{j_1...j_{n-1}}^{\infty}(\sigma_1 ,...\sigma_n )=lim_{L\to +\infty}
\Im_{j_1...j_{n-1}}(\sigma_1 L,...\sigma_n L)$$
satisfy the
same equations as conformal blocks;
\  Their  braiding  is
described by the quantum 6-j symbols
\ffk ,
\ref\alg{L. Alvarez-Gaume, C. Gomez  and G. Sierra, "Quantum group
interpretation of some conformal field theories", memorial
volume for Vadim Knizhnik, L. Brink, D. Friedan, A. Polyakov eds.,
 World Scientific (1989).},
\ref\mor{G. Moore and N. Seiberg,  Commun. Math. Phys. 123 (1989) 177.},
\ref\sau{G. Moore and N.Yu. Reshetikhin, Nucl. Phys. B328 (1989) 557.}.
Braiding properties do not uniquely specify conformal blocks .
For example,
the chiral correlation  functions of conformal
descendants have the same monodromic  properties as
correlators of primary fields. The braiding  will
uniquely determine  conformal blocks  if we
describe their singularities at coincident  points\ $\sigma_i\to\sigma_j$\ .
In the present case on can find the character of singularities
from the  explicit form of the  two point function\ \bvh .
In the limit\ $\beta\to\infty\,
(\Im m\beta\leq 0)$\ the function\ $g(\beta)$\ \nbv\
has the following asymptotic behavior:
\eqn\gdtr{g(\beta)=k^{\frac{1}{2}}\left(i\frac{\beta}{\pi\zeta}\right)^
{\frac{\alpha_+^2} {2}}(1+O(\frac{1}{\beta})).}
Hence the conformal dimension of
the field\ $\chi_a(\sigma)$\  is given by:
\eqn\gdfeer{\bar \Delta[\chi_a(\sigma)]=\bar \Delta_{(2,1)},}
where the standard notation\
\dot\
for the Kac spectrum is used
\eqn\dretys{\bar \Delta_{(l,l')}\equiv \frac{(\alpha_+l+\alpha_-l')^2
-(\alpha_++\alpha_-)^2}{4}.}
In this way  we
identify the functions \ $\Im_{j_1...j_{n-1}}^{\infty}(\sigma_1,...\sigma_n)$\
with the n-point conformal blocks of the fields \ $\Phi_{(2|1)}(\sigma)$\
\bpz.
The corresponding central charge is expressed  by the parameter\ $\xi$\
as follows:
\eqn\retsfd{c=1-\frac{(\alpha_+ +\alpha_-)^2}{6}.}

Similar arguments show that  the operator
\eqn\wer{\hat Z_a(\alpha)=
\exp\big(a\frac{\alpha}{2(\xi+1)}\big) Z_a(\alpha)}
can be regarded as  the field $\Phi_{(1|2)}$\
in the limit \ $\alpha\to\infty$\ .

In spite of the remarkable connection between operators \ \terass\  and
the fields\ $\Phi_{(2|1)},\Phi_{(1|2)}$\  ,
it is important to
understand their physical differences.
We consider the variable\ $\beta$\ as a rapidity
of physical excitations in massive models
and it cannot be identified with the holomorphic coordinate in the
2D CFT. From this point of view the discussed connection seems
to be a puzzle.

\subsec{Integral representation for form-factors in
SGM}

In order to get the integral representation for form-factors we
have to introduce an  analogue of the operators  \ $\Lambda_m(\alpha),
T(\alpha)$\ for  SGM.
It can be done by using the operators
\ $\hat Z_a(\alpha)$\ \wer .
The following formulas are a  generalization of \ \jhfx
\eqn\jh{T(\alpha)=\frac{(2 \cos\pi \nu)^{\frac{1}{2}}}{i}
\left[\matrix{\frac{1}{2}&\frac{1}{2}&0
\cr\frac{a}{2}&\frac{b}{2}&0}
\right]_{q'}
\hat Z'_a(\alpha+i\frac{\pi}{2})
\partial_{\alpha}\hat Z'_b(\alpha-i\frac{\pi}{2}),}
\eqn\hfgdt{\eqalign{\Lambda_m(\alpha)= \frac{i}{\eta'}
\left[\matrix{\frac{1}{2}&\frac{1}{2}&1\cr\frac{a}{2}&\frac{b}{2}&m}
\right]_{q'}\hat Z'_a(\alpha+i\frac{\pi}{2})
\hat Z'_b(\alpha-i\frac{\pi}{2}),}}
here\ $\nu\equiv\frac{1}{\xi+1}$\ .
Using this definition, one can obtain the bosonic representation
for \ $\Lambda_m(\alpha)$, which is necessary for evaluations of
form-factors.
\eqn\fgd{\eqalign{
&\Lambda_{1}(\alpha)=\tilde V'(\alpha),\cr
&\Lambda_0(\alpha)=
-\frac{i}{(2 \cos\pi \nu)^{\frac{1}{2}}}
\left[q'{\cal X}'\tilde V'(\alpha)-
q^{'-1}\tilde V'(\alpha){\cal X}'\right],\cr
&\Lambda_{-1} (\alpha)=-\frac{1}{2 \cos\pi\nu}\left[
q'{\cal X}^{'2}\tilde V'(\alpha)-(q'+q^{'-1}){\cal X}'\tilde
V'(\alpha){\cal X}'+
q^{'-1}\tilde V'(\alpha){\cal X}^{'2}\right].}}

As  in  the case of  \ $SU(2)$\ TM, the properties of the
operators \ $\Lambda_m(\alpha),
T(\alpha)$\ are  described by  Proposition 4 from  Sec. 7  .
So,  ${\cal F}_{a_1...a_n}^{m_1...m_k}(\alpha_1,...\alpha_k|
\beta_1,...\beta_n)$\ \hgdds\   will be  the generating functions for
form-factors in  SGM. The evaluations of  traces can be
done by
the technique discussed in Sec. 9.
I should make the
following remark only; The trace over the zero modes essentially
depends on  arithmetical properties of the interaction constant \ $\xi$\ .
The same phenomena take place when  conformal
blocks on  a torus are calculated
\ref\felders{ G. Felder, Nucl. Phys. B317 (1989) 215 .}.
To avoid  difficult problems connected with
a reducibility of the representation\ $\pi_Z$\ , we shall consider a
general case
when the parameter \ $\xi$\
is an irrational number greater then one.
Then  traces of the vertex operators\ \tdrsas\  are
represented by the formula\ \nvbc\ ,
where
the functions and constants have the following  forms:

\eqn\jdhk{\eqalign{&T_0(\{\alpha_i\},\{\delta_i\},\{\beta_i\},\{\gamma_i\})=
\delta_{k,p}\delta_{2n,r}\times \cr
&\ \ \ \ \ \ \times \exp\big(\frac{1}{2\xi}\big[\sum_{j=1}^n\beta_j
-2\sum_{j=1}^r\gamma_j\big]+\frac{1}{\xi+1}\big[\sum_{j=1}^p\delta_j
-\sum_{j=1}^k\alpha_j\big]\big) ,\cr
&G(\beta)= i\  {\cal C}_1 \sinh \frac{\beta}{2}
\exp\big[\int_{0}^{+\infty}\frac{d t} {t} \frac {\sinh^2t (1-i \frac{\beta}
{\pi}) }{\sinh 2 t \cosh t}\frac{\sinh t(\xi-1)}{\sinh t \xi}\big]\ ,\cr
&W(\beta)= -\frac{2}{\cosh \beta}
\exp\big[-2\int_{0}^{+\infty}\frac{d t} {t} \frac {\sinh^2t (1-i \frac{\beta}
{\pi}) }{\sinh 2 t }\frac{\sinh t(\xi-1)}{\sinh t \xi}\big]\ ,\cr
&\bar G(\beta)=-\frac{{\cal C}_2}{4}\ \xi\sinh\frac{ (\beta+i\pi)}{\xi}
 \sinh\beta\ , \cr
&\bar G'(\alpha)=-{\cal C}'_2\ \frac{\sinh\alpha}{(\xi+1)
\sinh\frac{\alpha+i\pi}{\xi+1}}\ , \cr
&U(\alpha)=i\ \sinh\frac{\alpha}{2}\ ,\cr
&\bar H(\alpha)=-\frac{2}{\cosh\alpha}\ , \cr
&{\cal C}_1=
\exp\big[-\int_{0}^{+\infty}\frac{d t} {t} \frac {\sinh^2\frac{t }
{2} }{\sinh 2 t \cosh t}\frac{\sinh t(\xi-1)}{\sinh t \xi} \big]\ , \cr
&{\cal C}_2=\exp\big[4\int_{0}^{+\infty}\frac{d t} {t} \frac {\sinh^2\frac{t }
{2} }{\sinh 2 t }\frac{\sinh t(\xi-1)}{\sinh t \xi} \big]\ ,\cr
&{\cal C}'_2=\exp\big[-4\int_{0}^{+\infty}\frac{d t}
{t} \frac {\sinh^2\frac{t } {2} }
{\sinh 2 t }\frac{\sinh t\xi}{\sinh t( \xi+1)} \big]\ .}}

Evaluations of   integrals which define the
generating  functions \ \hgdds\ is a problem complicated enough. I am going
to discus it in a separate publication.

\newsec{Conclusion}

At the end I wish to point out that
functions
\eqn\nvbb{{\cal F}^{m_1...m_k}(\alpha_1,...\alpha_k)=
Tr_{\pi_Z}\big[\exp(2\pi i K)
\Lambda_{m_k}(\alpha_{k})...\Lambda_{m_1}(\alpha_1)]}
are of special interest.
One can expect that they
and more general objects
$$Tr_{\pi_Z}\big[\exp(2\pi i K)
Z'_{a_n}(\alpha_n)...Z'_{a_1}(\alpha_1)]$$
represent   some vacuum  correlation functions
in  SGM.
The simplest of them have the following
explicit forms:
\eqn\hfgdf{\eqalign{Tr_{\pi_Z}\big[\exp(2\pi i K)
Z'_{a}&(\alpha_2)Z'_{b}(\alpha_1)]=
\cr
=&\C_{ab}\frac{\exp\frac{a\nu}{2}(\alpha_1-
\alpha_2-i\pi)}{4\nu\cos\pi\nu}
\frac{\sinh\nu(\alpha_1-\alpha_2+i\pi)}{\cosh\frac{ \alpha_1-\alpha_2}{2}}
\frac{G'(\alpha_1-\alpha_2)}{G'(-i\pi)} \ ,}}
where \ $G'(\alpha)$\ is given by\
\eqn\jfhhgp{G'(\alpha)=
\exp\big[\int_{0}^{+\infty}\frac{d t} {t} \frac {\sinh^2t (1-i \frac{\alpha}
{\pi}) }{\sinh 2 t \cosh t}\frac{\sinh t\xi}{\sinh t( \xi+1)}\big]\ ,}
and
\eqn\kfjgh{\eqalign{Tr_{\pi_Z}\big[\exp(2\pi i K)&
\Lambda_{m}(\alpha_2)\Lambda_n(\alpha_1)]=\cr
&=\delta_{m+n,0} (-1)^{m+1}
q^{'m}\frac{\sin 2\pi\nu}{16}
\frac{\theta(\alpha_1-\alpha_2 +i\pi)}{\sinh\nu(\alpha_1-\alpha_2+i\pi)}.}}
Fourier transformation of
the function\ $\theta(\alpha)$\  for \ $\xi>2 $ reads:
\eqn\hfgd{\theta(\alpha)=\int_{-\infty}^{+\infty}d t
\frac{ \tanh\frac{\pi t}{2\nu}\ \exp(i\alpha t)}
{\cos\frac{\pi}{2}(\nu-i t)\cos\frac{\pi}{2}(\nu+i t)
\cos\frac{\pi}{2}(3\nu-i t)
\cos\frac{\pi}{2}(3\nu+i t)}\ .}
Here \ $\nu\equiv\frac{1}{\xi+1}$\ .

It seems  important to clear up  a  physical meaning of these functions.

\centerline{}

\centerline{\bf Acknowledgments}

\centerline{}

I  would like to thank D. Brazhnikov, R. Chatterjee, V.A. Fateev,
S. Shatasvili,  S. Shenker, Al.B. Zamolodchikov
and especially A.B. Zamolodchikov
for helpful discussions.
This work  was  supported by grant DE-FG05-90ER40559.

\newsec{Appendix 1}

Here I give a  draft of the  proofs of  Propositions 1,2 from
Sec.5.

Let us begin with  Proposition 1.
The commutation relation
\eqn\jdhf{Z_+(\beta_1)Z_+(\beta_2)=S_{++}^{++}(\beta_1-\beta_2)
Z_+(\beta_2)Z_+(\beta_1)}
is evident from  the definition \ \rimaq\ .  To prove the formula
\eqn\jfdhf{Z_+(\beta_1)Z_-(\beta_2)=S_{+-}^{+-}(\beta_1-\beta_2)
Z_-(\beta_2)Z_+(\beta_1)+S_{-+}^{+-}(\beta_1-\beta_2)
Z_+(\beta_2)Z_-(\beta_1),}
we have to use the  relation:
\eqn\jfhf{\eqalign{{\cal X} V(\beta_2)V(\beta_1)=
&\frac{i\pi+\beta_1-\beta_2}{\beta_2-\beta_1}
V(\beta_2){\cal X}V(\beta_1)+\cr
&\frac{i\pi+\beta_2-\beta_1}{\beta_1-\beta_2}
S(\beta_2-\beta_1)V(\beta_1){\cal X}V(\beta_2)-V(\beta_2)V(\beta_1){\cal X}.}}

It may  be derived  in  the following way.
Consider the algebraic identity:
\eqn\lfdkfj{\eqalign{
(\beta_1-\beta_2)&(i\frac{\pi}{2}-\beta_1)(i\frac{\pi}{2}-\beta_2)+
(i\pi+\beta_1-\beta_2)(i\frac{\pi}{2}-\beta_1)(i\frac{\pi}{2}+\beta_2)=
\cr
&(i\pi+\beta_2-\beta_1)(i\frac{\pi}{2}+\beta_1)(i\frac{\pi}{2}-\beta_2)+
(\beta_2-\beta_1)(i\frac{\pi}{2}+\beta_1)(i\frac{\pi}{2}+\beta_2).}}
It is equivalent to the relation between the  functions\ $g(\beta)$\
and\ $w(\beta)$\ :
\eqn\pfdlfkg{\eqalign{w(\beta_2-\gamma)w(\beta_1-\gamma)g(\beta_1-\beta_2)=
\frac{i\pi+\beta_1-\beta_2}{\beta_2-\beta_1} &w(\gamma-\beta_2)
w(\beta_1-\gamma) g(\beta_1-\beta_2)+\cr
\frac{i\pi+\beta_2-\beta_1}{\beta_1-\beta_2}
S(\beta_2-\beta_1)w(\beta_2-\gamma) w(\gamma&-\beta_1)g(\beta_2-\beta_1)-\cr
w(\gamma&-\beta_2)w(\gamma-\beta_1)g(\beta_2-\beta_1).}}
Hence
\eqn\jfhfpo{\eqalign{\bar V(\gamma)V(\beta_2)V(\beta_1)&=
\frac{i\pi+\beta_1-\beta_2}{\beta_2-\beta_1} V(\beta_2)\bar V(\gamma)
V(\beta_1)+\cr&\frac{i\pi+\beta_2-\beta_1}
{\beta_1-\beta_2}S(\beta_2-\beta_1)V(\beta_1)\bar V(\gamma)
V(\beta_2)-V(\beta_2)V(\beta_1)\bar V(\gamma).}}
One can integrate both the  parts of this equation over
the variable \ $\gamma$\  and  get   \ \jfhf\  after  proper deformations
of the integration contours.

Using the formula \ \jfhf \  and the  definition\ \ters\ ,
it is easy to derive \ \jfdhf.
Other commutation relations for the Zamolodcikov-Faddeev algebra
can be obtained by  similar arguments.

Now let us prove  Proposition 2.
Its statement for the operator product
\eqn\mbn{Z_+(\beta_2)Z_+(\beta_1)=
\rho^2 g(\beta_1-\beta_2)\exp\big(-\frac{\beta_1+\beta_2}{2\xi}\big)
:V(\beta_2)V(\beta_1):}
follows from the explicit form of the function \ $g(\beta)$\ \yt\ .
Consider the operator product:
\eqn\hfg{\eqalign{Z_+(\beta_2)Z_-(\beta_1)=
\rho^2 \bar \rho&\eta^{-1}
g(\beta_1-\beta_2)\exp\big(\frac{\beta_1-\beta_2}{2\xi}\big)
\times\cr
\Biggr[q^{\frac{1}{2}}\int_{C}
\frac{d \gamma}{2\pi}& w(\gamma-\beta_1) w(\beta_2-\gamma)
:V(\beta_1) V(\beta_2)\bar V(\gamma):-\cr
-&q^{-\frac{1}{2}}\int_{C_1} \frac{d \gamma}{2\pi}
w(\gamma-\beta_1) w(\gamma-\beta_2)
:V(\beta_1) V(\beta_2)\bar V(\gamma):\Biggr]\ .}}
Here the integration contours are the same as  in the formula\ \fsrd\ .
There is one possibility of getting a singularity in the operator product
\ \hfg\  . It  appears when two integrand's poles clutch the
integration contour. Using the  form of the
function \ $w(\beta)$\ \hdbc\ , we can find that the second term in
\ \hfg\ is regular and the first one has a
simple pole  with the proper residue for
$\beta_2=\beta_1+i\pi.$

The analytical properties of the operator products
\ $Z_-(\beta_2)Z_+(\beta_1)$\ and\ $Z_-(\beta_2)Z_-(\beta_1)$\
can be investigated in a similar manner.

\newsec{Appendix 2}

In this Appendix I list the explicit expressions for the
functions and constants, which  describe the
operator products:
\eqn\reafrew{\eqalign{&V(\beta_2)V(\beta_1)=
\rho^2 g(\beta_1-\beta_2):V(\beta_2)
V(\beta_1):\ ,\cr
&\bar V(\gamma)V(\beta)=\rho\bar\rho
w(\beta-\gamma):\bar V(\gamma)V(\beta):\ ,\cr
&\bar V(\gamma_2)\bar V(\gamma_1)=
\bar\rho^2\bar g(\gamma_1-\gamma_2):\bar V(\gamma_2)\bar V(\gamma_1):\ ,\cr
&V'(\alpha_2)V'(\alpha_1)=\rho^{'2}g'(\alpha_1-\alpha_2)
:V'(\alpha_2)V'(\alpha_1):\ ,\cr
&\bar V'(\delta)V'(\alpha)=\rho'\bar\rho' w'(\alpha-\delta)
:\bar V'(\delta) V'(\alpha):\ ,\cr
&\bar V'(\delta_2)\bar V'(\delta_1)=\bar\rho^{'2}\bar g'(\delta_1-\delta_2)
:\bar V'(\delta_2)\bar V'(\delta_1):\ ,\cr
&V'(\alpha)V(\beta)=\rho\rho'h(\beta-\alpha):V(\beta)V'(\alpha):\ ,\cr
&\bar V'(\delta)V(\beta)=\rho\bar\rho'u(\beta-\delta)
:V(\beta)\bar V'(\delta):\ ,
\cr
&V'(\alpha)\bar V(\gamma)=\bar\rho\rho'u(\gamma-\alpha)
:\bar V(\gamma)V'(\alpha):\ ,\cr
&\bar V'(\delta)\bar V(\gamma)=\bar\rho\bar\rho'\bar h(\gamma-\delta)
:\bar V(\gamma) \bar V'(\delta): .}}

The functions\ $g,w,\bar g,g',w',\bar g',h,u,\bar h$\  have the
following forms for   \ $SU(2)$\ TM
\eqn\ytoi{\eqalign{&g(\beta)=
k^{\frac{1}{2}}\  \frac{\Gamma(\frac{1}{2}+\frac{i\beta}{2\pi})}
{\Gamma(\frac{i\beta}{2\pi})}\ ,
\cr
&w(\beta)=k^{-1}\  \frac {2 \pi}{i (\beta+i\frac{\pi}{2})}\ ,
\cr
&\bar g(\beta)=-k^2\  \frac{\beta (\beta+i \pi)}{4 \pi^2}\ ,
\cr
&g'(\alpha)=
k^{\frac{1}{2}}\  \frac{\Gamma(1+\frac{i\alpha}{2\pi})}
{\Gamma(\frac{1}{2}+\frac{i\alpha}{2\pi})}\ ,
\cr
&w'(\alpha)=k^{-1}\  \frac {2 \pi}{i (\alpha-i\frac{\pi}{2})} \ ,
\cr
&\bar g'(\alpha)=-k^2\  \frac{\alpha (\alpha-i \pi)}{4 \pi^2}\ ,\cr
&h(\beta)=k^{-\frac{1}{2}}\ \frac{\Gamma(\frac{1}{4}+\frac{i\beta}{2\pi})}
{\Gamma(\frac{3}{4}+\frac{i\beta}{2\pi})}\ ,
\cr
&u(\beta)=k\  \frac{i \beta}{2\pi}\ ,\cr
&\bar h(\beta)=-k^{-2}\  \frac{4\pi^2}{\beta^2+\frac{\pi^2}{4}}\ .}}
The constants \ $\rho,\bar \rho,\rho',\bar \rho' $\
are given by
\eqn\fghs{\eqalign{&\rho^2=i\big[\frac{k}{4\pi}]^{\frac{1}{2}}\ ,\ \
\bar\rho^2=-i \frac{k^2}{4\pi}\ ,\cr
&\rho^{'2}=\big[\frac{k}{\pi}\big]^{\frac{1}{2}}\ , \ \
\bar\rho'^2=i \frac{k^2}{4\pi}\ .}}
The constant \ $k$\ is  connected with the parameter of the ultraviolet
cut-off
\ $\epsilon$\ \lgjh\   as follows:
\eqn\gdfsasq{k=1-\exp(-2\pi\epsilon)\ .}
In the case of  \  $SU(2)$\ TM, the constants\ $\eta,\eta'$\
in the definitions of the  operators \ ${\cal X}, {\cal X}'$\
equal:
\eqn\bcvv{\eta=\big[-i \pi]^{\frac{1}{2}}\ , \ \ \
\eta'=\big[i \pi]^{\frac{1}{2}}\ .}

Now , let us consider  the case of  SGM.
The functions\ $g,w,\bar g,g',w',\bar g',h,u,\bar h$\  are given by:
\eqn\ytoisa{\eqalign{&g(\beta)=
\left[\frac{\kappa}{\Gamma(\frac{1}{\xi})}\right]^{\frac{1}{2}}
\frac{\Gamma(\frac{1}{\xi}+\frac{i\beta}{\pi\xi})}
{\Gamma(\frac{i\beta}{\pi\xi})}
\prod_{p=1}^{+\infty}\frac{\left[R_p(i\pi)R_p(0)\right]
^{\frac{1}{2}}}{R_p(\beta)}\ ,
\cr
&w(\beta)=
\kappa^{-1}\ \frac{\Gamma(-\frac{1}{2 \xi}+ \frac{i\beta}{\pi\xi})}
{\Gamma(1+\frac{1}{2\xi}+ \frac{i\beta}{\pi\xi})}\ ,
\cr
&\bar g(\beta)=\kappa^2\
\frac{i\beta}{\pi\xi}\frac{\Gamma(1+\frac{1}{\xi}+ \frac{i\beta}{\pi\xi})}
{\Gamma(-\frac{1}{\xi}+ \frac{i\beta}{\pi\xi})}\ ,
\cr
&g'(\alpha)=
\left[\kappa'\Gamma(\frac{1}{\xi+1})\right]^{\frac{1}{2}}
\frac{\Gamma(1+\frac{i\alpha}{\pi(\xi+1)})}
{\Gamma(\frac{1}{\xi+1}+\frac{i\alpha}{\pi(\xi+1)})}
\prod_{p=1}^{+\infty}\frac{R'_p(\alpha)}{\left[R'_p(i\pi)R'_p(0)\right]
^{\frac{1}{2}}}\ ,
\cr
&w'(\alpha)=
\kappa^{'-1}\ \frac{\Gamma(\frac{1}{2( \xi+1)}+ \frac{i\alpha}{\pi(\xi+1)})}
{\Gamma(1-\frac{1}{2(\xi+1)}+ \frac{i\alpha}{\pi(\xi+1)})}\ ,
\cr
&\bar g'(\alpha)=
\kappa^{'2}\
\frac{i\alpha}{\pi(\xi+1)}\frac{\Gamma(1-\frac{1}{ \xi+1}+
 \frac{i\alpha}{\pi(\xi+1)})}
{\Gamma(\frac{1}{\xi+1}+ \frac{\alpha}{i\pi(\xi+1)})}\ ,
\cr
&h(\beta)=k^{-\frac{1}{2}}\ \frac{\Gamma(\frac{1}{4}+\frac{i\beta}{2\pi})}
{\Gamma(\frac{3}{4}+\frac{i\beta}{2\pi})}\ ,
\cr
&u(\beta)=k\  \frac{ i\beta}{2\pi}\ ,\cr
&\bar h(\beta)=-k^{-2}\  \frac{4\pi^2}{\beta^2+\frac{\pi^2}{4}}.}}
Here \ $R_p(\beta)$\ and\ $R'_p(\alpha)$\ are  defined by
the equations\ \bvvc ,\ \bvvcre\ .
The constants \ $\rho,\bar \rho,\rho',\bar \rho' $\
have the following  values  for  SGM:
\eqn\bsare{\eqalign{&\rho^2=\frac{i}{\pi\xi}\left[\kappa \Gamma(\frac{1}{\xi})
\right]^{\frac{1}{2}}\prod_{p=1}^{+\infty}
\left[\frac{R_p(i\pi)}{R_p(0)}\right]
^{\frac{1}{2}}\ ,
\cr
&\bar \rho^2=\kappa^2\
\frac{i}{\pi\xi}\frac{\Gamma(1+\frac{1}{\xi})}
{\Gamma(-\frac{1}{\xi})}\ ,
\cr
&\rho^{'2}=\left[\frac{\kappa'}{\Gamma(\frac{1}{\xi+1})}\right]^{\frac{1}{2}}
\prod_{p=1}^{+\infty}\left[\frac{R'_p(0)}{R'_p(i\pi)}\right]
^{\frac{1}{2}}\ ,\cr
&\bar \rho^{'2}=
\kappa^{'2}\
\frac{i}{\pi(\xi+1)}\frac{\Gamma(1-\frac{1}{ \xi+1})}
{\Gamma(\frac{1}{\xi+1})} \ .}}
The constants \ $\kappa,\kappa',k$\ are connected with the parameter of the
ultraviolet cut-off\ $\epsilon$\  as follows:
\eqn\hdhgf{\kappa=\left[1-\exp(-\pi\xi\epsilon)\right]^{\frac{\xi+1}{\xi}}
\ ,\ \
\kappa'=\left[1-\exp(-\pi(\xi+1)\epsilon)\right]^{\frac{\xi}{\xi+1}}\ ,\ \
k=1-\exp(-2\pi\epsilon)\ .}
In the case of  SGM  the constants\ $\eta,\eta'$\ equal
\eqn\bcvv{\eta=\left[-i\xi\sin\frac{\pi}{\xi}\right]^{\frac{1}{2}}\ , \ \ \
\eta'=\left[i(\xi+1)\sin\frac{\pi}{\xi+1}\right]^{\frac{1}{2}}\ .}

\newsec{Appendix 3}

In this Appendix the structure functions for the algebra\ \gdfs  -\bcvd\
are listed in the case of SGM.
The structure functions for \ $SU(2)$\ TM can be obtained from
the presented ones
by taking  a limit\ $\xi\to\infty$\ .

The matrix \ ${\cal R}_{ab}^{cd}$\ is the
R-matrix for the  19-vertex model
\ref\fatee{V.A. Fateev and A.B. Zamolodchikov,
Sov. Journ. Nucl. Phys. 32 (1980) 298.}.
It can be represented as follows:
\eqn\cvxz{\eqalign
{&{\cal R}_{m_1 m_2}^{m_3 m_4}(\alpha)=
\sum_{\scriptstyle j=0,1,2\atop\scriptstyle
-j\le m\le j}d_j(\alpha)
\left[\matrix{1&1&j\cr
m_1&m_2&m}\right]_{q'}
\left[\matrix{1 &1 &j\cr
m_4&m_3&m}\right]_{q'} .}}
Here
\eqn\reqwe{\eqalign{&d_0(\alpha)=\frac{[\alpha-2i\pi]}{[\alpha+2i\pi]}\ ,\cr
&d_1(\alpha)=-\frac{[\alpha+i\pi]
[\alpha-2i\pi]}{[\alpha-i\pi]
[\alpha+2i\pi]}\ ,\cr
&d_2(\alpha)=\frac{[\alpha+i\pi] }{[\alpha-i\pi]}\ ,}}
and
\eqn\bgft{[x]\equiv\sinh\frac{x}{\xi+1} .}
The structure functions \ ${\cal A },{\cal B },{\cal C }$\ read
\eqn\gdfrs{\eqalign
{&{\cal A}_{m m_1m_2}(\alpha)=
\frac{[i\pi][ 4i\pi]^{\frac{1}{2}}}
{(\xi+1)^{\frac{1}{2}}[ \alpha-i\pi][\alpha+ 2i\pi]}
\left[\matrix{1&1&1\cr
m_1&m_2&m}\right]_{q'} \ ,}}
\eqn\hfgdop{\eqalign
{&{\cal A}_{m_1m_2}(\alpha)=
\frac{([i\pi ]^3[3i\pi] )^{ \frac{1}{2}}
[2\alpha]}{(\xi+1)[ \alpha-i\pi]
[\alpha][\alpha+i\pi][\alpha+ 2i\pi]}
\left[\matrix{1&1&0\cr
m_1&m_2&0}\right]_{q'}\ ,}}
\eqn\dfsaa{
{\cal A}(\alpha)=\frac{[ i\pi]^2 [ 2i\pi] }
{(\xi+1)[ \alpha-i\pi]
[\alpha][\alpha+i\pi][\alpha+ 2i\pi]}\ ,}
\eqn\dfsaasa{
{\cal B}(\alpha)=-\frac{[ i\pi]^2 [2\alpha+ 2i\pi] }
{(\xi+1)[ \alpha-i\pi]
[\alpha][\alpha+i\pi][\alpha+ 2i\pi]}\ ,}
\eqn\gtffs{{\cal C}(\alpha)=-\frac{ [i\pi ]^3[\alpha+3 i\pi]
[2\alpha]}{(\xi+1)[ 2i\pi]
[\alpha-i\pi]
[ \alpha][\alpha+i\pi]^2[\alpha+ 2i\pi]}+
(\xi+1)[i\pi]\p_{\alpha}^2\ln R(\alpha)
\ ,}
where  \ $R(\alpha)$\ is defined by the equation\ \bvvcre .

\listrefs

\end